\documentclass[journal]{IEEEtran}
\usepackage{times,amsmath,amsfonts,amssymb,amsthm}
\usepackage{cite}
\usepackage{graphicx,subfigure,psfrag}
\usepackage{algorithm}%,algorithmic}
\usepackage{booktabs}
\usepackage{multirow}
\usepackage{multicol}
\usepackage{colortbl}
\usepackage{color}
\usepackage{array}
\usepackage{algpseudocode}
\def\abf{{\bf a}}
\def\bbf{{\bf b}}
\def\cbf{{\bf c}}
\def\dbf{{\bf d}}

\def\gbf{{\bf g}}

\def\pbf{{\bf p}}

\def\xbf{{\bf x}}

\def\xbf{{\bf x}}

\def\Abf{{\bf A}}

\def\Ac{{\cal A}}

\def\Kc{{\cal K}}

\def\Nc{{\cal N}}
\def\Oc{{\cal O}}

\def\Sc{{\cal S}}

\def\Wc{{\cal W}}

\def\eg{{\it e.g.,\ \/}}

\def\ie{{\it i.e.,\ \/}}
\def\nn{\nonumber}

\newcommand{\tabincell}[2]{\begin{tabular}{@{}#1@{}}#2\end{tabular}}
\makeatletter
\newcommand{\thickhline}{%
    \noalign {\ifnum 0=`}\fi \hrule height 1pt
    \futurelet \reserved@a \@xhline
}

\theoremstyle{definition}
\newtheorem{lemma}{Lemma}
\newtheorem{theorem}{Theorem}
\newtheorem{Remark}{Remark}
\newtheorem{proposition}{Proposition}
\newtheorem{myDef}{Definition}
\newtheorem{corollary}{Corollary}

\begin{document}

\title{Fundamental Structure of Optimal Cache Placement for Coded Caching with  Nonuniform Demands}

\author{Yong Deng, \IEEEmembership{Student Member, IEEE}  and Min~Dong, \IEEEmembership{Senior Member, IEEE}\thanks{The authors are with the Department of Electrical, Computer and Software Engineering, Ontario Tech University, Oshawa, ON L1G0C5, Canada (email: \{yong.deng, min.dong\}@ontariotechu.ca). Preliminary results in this work were presented in \cite{Yong&Dong:Asilomar19}.}}

%\IEEEauthorblockA{Dept. of Electrical, Computer and Software Engineering\\ %University of Ontario Institute of Technology, Ontario, Canada}
%\IEEEauthorblockA{Email: \{yong.deng, min.dong\}@uoit.ca}

\maketitle

\allowdisplaybreaks

\begin{abstract}
This paper studies the caching system of  multiple cache-enabled users with random demands. Under nonuniform file popularity, we thoroughly characterize the optimal uncoded cache placement structure for the coded caching scheme (CCS). Formulating the cache placement as an optimization problem to minimize the average delivery rate, we identify the file group structure in the optimal solution. We show that, regardless of  the file popularity distribution, there are \emph{at most three file groups} in the optimal cache placement{, where files within a group have the same cache placement}. We further characterize the complete structure of the optimal cache placement and obtain the closed-form solution in each of the three file group structures. A simple algorithm is developed to obtain the final optimal cache placement by comparing a set of candidate closed-form solutions  computed in parallel. We provide insight into the file groups formed by the optimal cache placement. The optimal placement solution also indicates that coding between file groups may be explored during delivery, in contrast to the existing suboptimal file grouping schemes. Using the file group structure in the optimal cache placement for the CCS, we propose a new information-theoretic converse bound for coded caching that is tighter than the existing best one. Moreover, we characterize the file subpacketization in the  CCS with the optimal cache placement solution and show that the maximum subpacketization level in the worst case scales as $\Oc(2^K/\sqrt{K})$ for $K$ users.
%Simulations verify   the optimal placement solution produced by our algorithm and the file group structure in it. The optimal CCS outperforms existing schemes, especially for  smaller cache size. 
%The  tighter converse bound and file subpacketization level under the optimal cache placement are also demonstrated. 
\end{abstract}
%
%%
%% Note that keywords are not normally used for peerreview papers.

\section{Introduction}
\label{sec:introduction}

Future wireless networks face  rapid data  traffic growth and  increasing demands for timely content delivery. Caching has emerged as a promising technology to  address these pressing issues~\cite{Bastug&etal:COMMag14,Wang&etal:COMMag14,paschos2018role}.
By storing data in distributed network storage resources near  base stations or user devices, cache-aided systems alleviate the increasingly intensive traffic in wireless networks to meet low latency requirements.
Conventional uncoded caching can improve the hit rate~\cite{baev2008approximation,borst2010distributed,tan2013optimal,shanmugam2013femtocaching} but is not efficient when there are multiple cache-aided devices.
Coded caching has been recently introduced in the seminal work~\cite{Maddah-Ali&Niesen:TIT2014}, where a {coded caching scheme} has been proposed. It combines a carefully designed cache placement of uncoded contents and a coded multicast delivery strategy to explore the caching gain. For convenience, we refer to this {coded caching scheme} as  \emph{the CCS} in the rest of this paper. By exploring both global and local caching gain, the CCS was shown to be able to serve an infinite number of users simultaneously with finite resources \cite{Maddah-Ali&Niesen:TIT2014}. Since then, coded caching has drawn considerable attention, with extensions of the CCS to the decentralized scenario \cite{Niesen&Maddah-Ali:TIT2015}, transmitter caching in mobile edge networks \cite{Sengupta&etal:TIT17,Maddah-Ali&etal:ISIT}, user caching in device-to-device networks\cite{ji2016fundamental}, and transmitter-and-receiver caching in wireless interference networks \cite{Xu&etal:TIT17}.

A key design issue in coded caching is  cache placement. An effective cache placement scheme maximizes caching gain and minimizes the transmission load  in the network in the content delivery phase  (\ie the delivery rate). The works mentioned above all assume uniform file popularity under homogenous demands, for which a symmetric cache placement strategy (\ie identical cache placement for all files) is optimal \cite{Daniel&Yu:TIT19}.
In the more general scenario of   files with heterogeneous   demands leading to nonuniform file popularity, the cache placement may be different among files, complicating both caching design and analysis.
There is a fundamental question on whether to distinguish files of different popularities and to what extent.
On the one hand, different cache placements for files with distinct popularities may help capture the  difference in demands to improve caching efficiency.
On the other hand, depending on the degree of  difference, ignoring this difference  in file popularity and simply using the symmetric cache placement may be a good tradeoff between  performance gain and implementation complexity.

Several recent works have   considered the cache placement design for the CCS  under nonuniform file popularity  \cite{Niesen&Maddah-Ali:TIT2017,hachem2017coded,Ji&Order:TIT17,Zhang&Coded:TIT18}, where
a typical method  is to construct a cache placement scheme  and  bound its performance.
For complexity reduction, file grouping is commonly used as a tractable  method for the cache placement design. 
It was first proposed in \cite{Niesen&Maddah-Ali:TIT2017},  which  divides files into groups based on their popularities and allocates chunks of cache  to different groups. Files within each file group are treated the same with  identical cache placement.
 Following this, different methods to partition files into   file groups have  been proposed  \cite{Ji&Order:TIT17,hachem2017coded,Zhang&Coded:TIT18}. 
These
existing studies show that file grouping is an effective method
to handle nonuniform file popularity for cache placement.
However, the  file grouping methods used in these existing schemes  are all somewhat  heuristic, with two file groups  typically considered to separate the most popular files from the remaining ones. The optimal cache placement and its relation
to file grouping remain unknown.
Different from the method of construction, the optimization approach was adopted in   \cite{Daniel&Yu:TIT19,Jin&Cui:Arxiv2017} to  formulate the  cache placement into an optimization problem to find the  solution. %A property of the optimal cache placement that   more %cache is allocated to a file with higher popularity is shown \cite{Jin&Cui:Arxiv2017}.
Both works have focused on developing numerical methods to solve the optimization problem.  However, the numerical  results cannot provide insights  into the optimal cache placement.

Indeed,  characterizing the optimal cache placement structure may bring us a deeper understanding of  the effect of nonuniform file popularity on the caching gain offered by the CCS. Furthermore, in the cache placement for the CCS, each file is partitioned into  subfiles to be stored at different sets of users. The number of required subfiles may potentially grow exponentially with the number of users.
This could prevent the practical use of the CCS for files with finite sizes and limit the caching gain that can be achieved. There have been  studies on the tradeoff between the subpacketization level and the coded caching gain by the CCS under uniform file popularity \cite{shanmugam2016finite,yan2017placement,cheng2017coded,tang2017low,shanmugam2017coded,lampiris2018adding}. The analysis of subpacketization is more challenging for nonuniform file popularity and, therefore, scarce in the literature, as different files may be partitioned in different ways. Obtaining the optimal cache placement structure will help characterize the    file subpacketization in the CCS to understand the  practical limits and make an appropriate  tradeoff between the subpacketization level and coded
caching gain for the CCS.   %File grouping is also used to derive  information theoretic converse bounds %for coded caching. along with a genie-based method, various information %theoretic converses have been developed~\cite{Niesen&Maddah-Ali:TIT2017,Ji&Order:TIT17,Zhang&Coded:TIT18,hachem2017coded}.

\subsection{Contributions}
In this paper, we  characterize the optimal cache placement for the CCS under nonuniform file popularity.  We obtain the optimal cache placement structure  and establish its connection to file group structure under arbitrary file popularity distribution and cache size. 

{Different from the construction method adopted in many existing works, we use the optimization framework to formulate the  uncoded cache placement problem  to minimize the average rate in the coded content delivery phase. The optimization problem is formulated to find the optimal cache placement  in a broad family of  centralized and decentralized placement schemes.  Exploring several properties of the optimization problem, we  reformulate the problem into a simplified yet equivalent linear programming (LP) problem.
We identify the inherent file group structure  in the optimal cache placement by analyzing the structure and the optimality conditions of the reformulated problem.} In particular, we show that there are \emph{at most three file groups} in the
optimal cache placement regardless of the file popularity distribution, with files in each group having an identical cache placement. Each  possible file group structure has a unique cache placement pattern. {With further  in-depth analysis of these patterns and caching constraints}, we characterize
the complete structure of the optimal cache placement and obtain the closed-form placement solution for  each of the three possible file group structures. In particular,  we show that each  file to be cached is partitioned into subfiles of at most two different sizes. Following these, we
 develop a simple and efficient algorithm to obtain the final optimal cache placement, which only requires computing a set of candidate solutions in closed-form  in parallel. 

{The result of at most three file groups in the optimal cache placement, regardless of file popularity distribution, is somewhat surprising. We provide insight into the above file grouping results. Despite different file popularities, the cache placement strategy  only distinguishes files as ``most popular," ``moderately popular," or ``non-popular," and based on these categories to determine  whether to cache the entire, a portion, or none of a file among users (see detailed discussion in Section \ref{subsubsec:opt_placement}). The files are mapped to one of these three categories  to form  file groups.   The optimal placement may have one to three  file groups, depending on the file popularity distribution and the ratio of global cache size to the database size. We point out that although two file groups have been considered in the existing decentralized caching schemes for the CCS \cite{Ji&Order:TIT17,Zhang&Coded:TIT18},  there is no existing  scheme that considers   either three file groups for  coded caching or coding between file groups. Our result shows that there can be three file groups in the optimal cache placement, and the coding opportunity between file groups may be explored during coded content delivery. 
%Although the result of at most three file groups, regardless of the number of files, file popularities, and cache size,  is surprising, we provide insight into the result in terms of popularity categories, and give intuition of the optimal placement in each category.
}

 The file group structure in the optimal cache placement solution {for the CCS} enables us to obtain a new information-theoretic lower bound on the average rate for any caching scheme under nonuniform file popularity.
It is derived by applying the optimal file group   structure obtained for the CCS to the genie-aided construction method~\cite{Ji&Order:TIT17,Zhang&Coded:TIT18}.
Our  lower bound is tighter than  the existing best one \cite{Zhang&Coded:TIT18}. This improvement  shows that the file groups resulting from the optimal cache placement provide a better indication of  the popular file group than the existing  methods suggest. %~\cite{Ji&Order:TIT17,Zhang&Coded:TIT18}.

 Based on the structure of the optimal cache placement, we are able to further characterize the  file subpacketization in the cache placement for the CCS.      We derive the maximum subpacketization level in the worst case and show that it scales as  $\Oc(2^K/\sqrt{K})$ for $K$ users.
Both analysis and simulation show that the general subpacketization can be much smaller than this
 upper bound.

 The optimal cache placement structure and the optimal solution obtained by our algorithm are verified and demonstrated through simulation. The optimal cache placement  outperforms other existing  schemes for the CCS. The performance gap is larger when the cache size is smaller, demonstrating that a better cache placement strategy is more critical to maximize the caching gain. The simulation also shows that the proposed lower bound is tighter than the existing ones for various system configurations. Finally, the subpacketization level for the optimal CCS and the impact of cache size on it are studied in simulation.

\subsection{Related Works}
\begin{table*}[t]
\renewcommand{\arraystretch}{1.05}
\centering
\caption{Comparison with existing cache placement schemes for the CCS}
\resizebox{15.55cm}{!}{
\begin{tabular}{|c|c|c|c|}
\hline
\multicolumn{1}{|l|}{} & Approach                                 & File grouping strategy   & Cache placement strategy \\ \hline\hline
[16]                   & Proposed a suboptimal scheme              & Multiple file groups     & Decentralized      \\ \hline
[17],[18], [19]        & Proposed a suboptimal scheme              & One or two file groups          & Decentralized      \\ \hline
\cite{Wang2019Optimization}               & Via optimization,  (suboptimal) numerical methods     & N/A                      & Decentralized      \\ \hline
[15], [20]             & Via optimization, numerical method     & N/A                      & Centralized        \\ \hline
Our work               & Via optimization, closed-form optimal solution & Optimal file groups & Centralized        \\ \hline
\end{tabular}
}
\label{table:compare}
\end{table*}

 The CCS has been studied  in many works for various system scenarios to understand the fundamental limit of coded caching~\cite{Maddah-Ali&Niesen:TIT2014,Niesen&Maddah-Ali:TIT2015,Sengupta&etal:TIT17,Maddah-Ali&etal:ISIT,ji2016fundamental,Xu&etal:TIT17,Daniel&Yu:TIT19}. In these works, the  cache placement  for the CCS has been studied for the peak delivery rate under uniform file popularity\footnote{For uniform file popularity, it can be shown that the peak rate and average rate are identical for the CCS.}, where the optimal cache placement in this case is the same for all files~\cite{Maddah-Ali&Niesen:TIT2014,Daniel&Yu:TIT19}. The cache placement  under nonuniform file popularity  has been investigated in~\cite{Niesen&Maddah-Ali:TIT2017,hachem2017coded,Ji&Order:TIT17,Zhang&Coded:TIT18}.  It was first studied in~\cite{Niesen&Maddah-Ali:TIT2017}, where a file grouping strategy {independent of the number of users $K$} was proposed to reduce the design complexity by treating files in each group to be the same and using the symmetric decentralized CCS for each group.
Following this, {by incorporating the knowledge of  $K$ in the file grouping design, several  suboptimal file grouping schemes have  been proposed to lower the average delivery rate}~\cite{hachem2017coded,Ji&Order:TIT17,Zhang&Coded:TIT18}.
In~\cite{hachem2017coded}, a  specific multi-level file popularity  model  is considered, where the number of files at each level and the number of users requesting the files at each level are fixed. Under this model, a caching scheme using two file groups was proposed and shown to be order-optimal depending on the number of levels.
With a more general file popularly distribution, a simple RLFU-GCC scheme was proposed in \cite{Ji&Order:TIT17}, which splits files into two file groups, with one containing the most popular files and allocated  the entire cache.
The performance of this scheme was shown to be order-optimal for the Zipf distribution.
As an extension to an arbitrary popularity distribution, a mixed caching strategy was proposed~\cite{Zhang&Coded:TIT18} by adding a choice of  an uncoded caching scheme to the above two-file-group caching scheme. This added scheme has three file groups for cache placement and uses uncoded delivery. All the above works \cite{Niesen&Maddah-Ali:TIT2017,hachem2017coded,Ji&Order:TIT17,Zhang&Coded:TIT18} use decentralized CCS  for each file group, and there is no coding opportunity between the file groups  in these schemes.
Different from the above approaches, the optimization framework is considered to find the optimal  cache placement for the CCS under nonuniform file popularity in the centralized scenario \cite{Daniel&Yu:TIT19,Jin&Cui:Arxiv2017} {and   the decentralized setting \cite{Wang2019Optimization}}.\footnote{From the optimization perspective, the decentralized cache placement problem is a subproblem of the  cache placement optimization problem in the centralized scenario. In other words, any decentralized cache placement is a feasible point of the centralized  cache placement optimization problem.} Numerical methods are resorted to solve these problems, which  cannot be used to characterize the optimal cache placement.
The optimal cache placement for the CCS under arbitrary  file popularity distribution and its relationship with file grouping remains unknown. {We summarize the differences between our work and the above mentioned existing works for the CCS in Table~\ref{table:compare}.}

 For understanding the  fundamental limit of coded caching, information-theoretic converse bounds are developed in the literature.  The lower bounds on the peak and average rates for files with uniform  popularity have been developed and improved by several works ~\cite{Maddah-Ali&Niesen:TIT2014,Niesen&Maddah-Ali:TIT2015,Wang2016Anew,Wang2018Improved}. For nonuniform file popularity, different lower bounds on the average rate have been developed to demonstrate the performance of the proposed file grouping based caching schemes~\cite{Niesen&Maddah-Ali:TIT2017,Ji&Order:TIT17,Zhang&Coded:TIT18,hachem2017coded}.
A lower bound was first developed in~\cite{Niesen&Maddah-Ali:TIT2017}, where  a genie-based method was used to compute the sum  peak delivery rates of multiple file groups that are heuristically partitioned.
The number of file groups depends on popularity distribution, and  the bound is generally loose.
The genie-based method is commonly used to obtain the lower bounds~\cite{Ji&Order:TIT17,Zhang&Coded:TIT18,hachem2017coded}. It constructs a virtual system where only a group of most popular files need to be delivered to the users via the shared link, and these popular files are treated equally. The group of most popular files is {formed either heuristically or through a suboptimal method}, resulting in different tightness of the lower bound. In~\cite{Ji&Order:TIT17},  focusing on the Zipf distribution of file popularity, the authors {proposed a method to determine} the group of most popular files for different Zipf parameters, and a lower bound on the average rate is developed using the peak delivery rate  in this  file group.
A lower bound for an arbitrary file popularity distribution  was obtained in  \cite{Zhang&Coded:TIT18} by categorizing the most popular files via a different strategy.
Furthermore,  a file merging process was proposed to  tighten the bound further by including some moderately popular files into the  group of most popular files. From these existing studies,  the proposed file grouping strategies  appear to have a strong influence on the tightness of the lower bound.
In this work, we show that the file group structure in the optimal cache placement would lead to a tighter lower bound.

File subpacketization in the cache placement has been studied in~\cite{shanmugam2016finite,yan2017placement,cheng2017coded,tang2017low} for uniform file popularity, where different  methods were proposed to reduce the subpacketization level  in the cache placement with a higher delivery rate as a tradeoff. The Pareto-optimal coded caching schemes that characterize the tradeoff between the high subpacketization level and the rate were provided in~\cite{cheng2017coded}.
The cache placement of the CCS~given in \cite{Maddah-Ali&Niesen:TIT2014} for specific cache sizes is proved to be both optimal~\cite{yan2017placement} and Pareto-optimal~\cite{cheng2017coded}  in achieving the highest cache gain with the minimum subpacketization level.
In \cite{shanmugam2017coded}, the existence of coded caching schemes with the linear growth of the subpacketization level for a  large number of users is shown.
In~\cite{lampiris2018adding},   using multiple antennas is suggested to reduce the subpacketization level.
The problem under nonuniform file popularity is much more complicated, and the study is scarce. In\cite{Jin&Cui:Arxiv2018},  a cache placement optimization problem is considered that uses the subpacketization level as a constraint. The influence of the subpacketization level on the average rate was explored through numerical simulations. Unfortunately,
the simulation approach is not able to provide insights into the subpacketization feature in the optimal cache placement solution.

Besides nonuniform file popularity, other types of nonuniformity have also been considered in  the coded caching design, including  file sizes \cite{Daniel&Yu:TIT19,Zhang2019Closing},   cache sizes  \cite{Yang2018Coded,Ibrahim2019Coded}, and link qualities \cite{Cacciapuoti2016speeding,Amiri2018Cache_aided,Cao2019coded}. In addition, a modified CCS has been proposed recently with an improved delivery strategy that results in a reduced delivery rate than the original CCS  \cite{Yu&Maddah-Ali:TIT2018}. The complication in the delivery strategy further complicates the analysis of this caching scheme. In this paper, we focus on the characterization of the optimal caching solution for the CCS to provide insights into the effect of cache placement on the caching gain.
%All the above combinatorial design based works are proposed only for uniform file popularity and specific memory sizes.Assuming nonuniform popularities and arbitrary memory sizes, \cite{Jin&Cui:Arxiv2018} built an optimization problem that uses the subpacketization level as a constraint, aiming to minimize the average rate based on a modified CCS. It uses numerical method to explore the influence of subpacketization level on the average rate. However, this method does not provide any insight into the file splitting method of the optimal cache placement solution.

\subsection{Organization and Notations}
The rest of the paper is organized as follows. In Section~\ref{sec:model}, we introduce the system model and describe the cache placement problem for the CCS. In Section~\ref{sec:problem}, we formulate the  cache placement optimization problem and transform it into a simpler form. In Section~\ref{sec:Placement}, we identify all possible file group structures under the optimal cache placement,  present the optimal cache placement solution in each  case, and provide a simple  algorithm to obtain the final optimal solution. In Section~\ref{sec:bound},  we propose a converse bound for  general coded caching   tighter than existing bounds.   In Section~\ref{sec:subpacket}, we derive an upper bound on  the subpacketization level for the CCS under the optimal cache placement. Simulation is provided in Section~\ref{sec:simu} to verify our results and demonstrate the performance. The conclusion is provided in Section~\ref{sec:conclusion}.

\emph{Notations}: The cardinality of set $\Sc$ is denoted by $|\Sc|$, and the size of file $W$ is denoted by $|W|$. The bitwise "XOR" operation between two subfiles is denoted by 
 $\oplus$. Notations $\lfloor\cdot\rfloor$ and $\lceil\cdot\rceil$ denote the floor and ceiling functions, respectively. Notation  $\abf \succcurlyeq {\bf 0}$ means element-wise  non-negative in vector $\abf$.
We extend the definition of ${K \choose l}$ and define ${K \choose l}=0$,  for  $l<0$ or $l>K$.

\section{System Model and Problem Setup}\label{sec:model}
\subsection{System Model}
\begin{figure}
  \centering
  \includegraphics[scale=0.4]{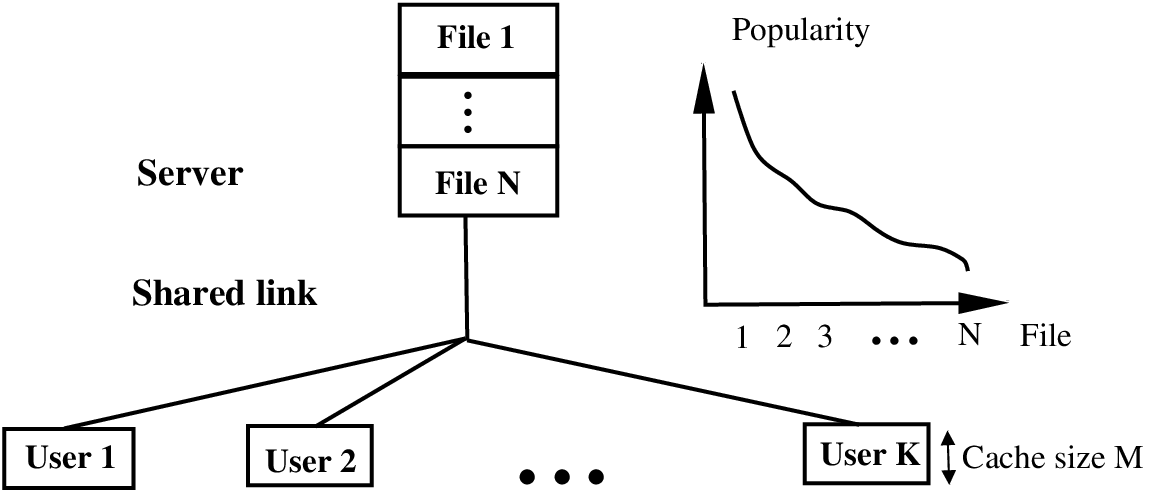}
  \caption{An example of cache-aided systems, where end users are connected
to the central service provider through a shared link. Each user has a local cache to alleviate the burden of the shared link. The files in the server have nonuniform popularities.} \label{fig:sys_mod}
\end{figure}

Consider a cache-aided transmission system with a server connecting to $K$ users, each with a local cache, over a shared error-free link, as shown in Fig.~\ref{fig:sys_mod}. The server has a database consisting of $N$ files, $\{W_1,\ldots,W_N\}$. Each file $W_n$ is of size $F$ bits and is requested with probability $p_n$. Let $\pbf=[p_1,\ldots,p_N]$ denote the popularity distribution of all $N$ files, where $\sum_{n=1}^{N}p_n=1$.
Without  loss of the generality, we label files according to  the decreasing order of their popularities: $p_1\geq p_2\geq \cdots\geq p_N$.
Each user $k$  has a local cache of capacity $MF$ bits, which is referred to as cache size $M$ (normalized by the file size), where  $M$ is a real number and $M\in[0,N]$.
Denote the file and user index sets by $\Nc\triangleq\{1,\ldots,N\}$ and $\Kc\triangleq\{1,\ldots,K\}$, respectively.

The coded caching operates in two phases: the cache placement phase and the content delivery phase.
In the cache placement phase, a portion of uncoded file contents from $\{W_1,\ldots,W_N\}$ are placed in each user $k$'s local cache, according to a cache placement scheme.
The cached content at user $k$ is described by a caching function $\phi_{k}(\cdot)$ of $N$ files as $Z_{k}\triangleq\phi_{k}(W_1,\ldots,W_N)$. 
During data transmission, each user $k$  independently requests a file with index $d_k$  from the server.
Let $\dbf\triangleq[d_1,\ldots,d_K]$ denote the demand vector of all $K$ users. In the content delivery phase, based on the demand vector $\dbf$ and the cached contents at users, the server generates coded messages of uncached portions of requested files and sends them to the users.
The generated codeword can be described by an encoding function  $\psi_{\dbf}(\cdot)$ of the $N$ files   for demand $\dbf$  as  $X_{\dbf}=\psi_{\dbf}(W_1,\ldots,W_N)$. Upon receiving the codeword, each user $k$ applies a decoding function $\varphi_k(\cdot)$ to reconstruct its (estimated) requested file $\hat{W}_k$ from the received codeword and its cached content  as $\hat{W}_k\triangleq\varphi_k(X_{\dbf},Z_{k})$. A valid coded caching scheme requires that  each user $k$ is able to reconstruct its requested file,  ${\hat W}_{\dbf,k}=W_{d_k}$, $k\in \Kc$, for any demand $\dbf$,  over an error-free link.

\subsection{Cache Placement Problem Construction}\label{III.A}
The cache placement is a crucial design issue in  coded caching. Among existing studies for the CCS, a common approach  is to propose a cache placement scheme, construct a lower bound on the minimum data rate, and evaluate the proposed scheme by comparing its performance  with the lower bound. In this work, we use an optimization approach for the cache placement design for the CCS.   Through construction,  we formulate the cache placement problem  into a design optimization problem.

\subsubsection{Cache placement}

For $K$ users, there are $2^K$ user subsets in $\Kc$, with subset sizes ranging from $0$ to $K$. Denote $\Kc_0  \triangleq\Kc \cup\{0\}$.  Among all the user subsets, there are $\binom{K}{l}$ different user subsets with the same size $l\in \Kc_0$ ($l=0$ corresponds to the empty subset $\emptyset$ in $\Kc$).
%Let $\Sc^{l}_{i}$ denote user subset $i$ with $|\Sc^{l}_i|=l$, for $i=1,\ldots,{K \choose l}$.
They form a cache subgroup that contains all user subsets of size $l$,  defined as $\Ac^l\triangleq\{\Sc: |\Sc|=l,\ \Sc\subseteq \Kc\}$ with $|\Ac^l|=\binom{K}{l}$, for $l\in \Kc_0$. For the $N$ files, partition each file $W_n$  into $2^K$ non-overlapping subfiles, one for each unique user subset $\Sc\subseteq\Kc $, denoted by
 $W_{n,\Sc}$ (it can be $\emptyset$). Each user $k \in \Sc$ stores subfile $W_{n,\Sc}$ in its local cache (for $\Sc=\emptyset$, subfile $W_{n,\emptyset}$   is not cached to any user, but  only  kept in the server). For any caching scheme, each  file should be  reconstructed  by combining all its subfiles. Thus, we have the file partitioning constraint
 \begin{align}\label{Constraint1}
   \sum_{l=0}^{K}\sum_{\Sc\in \Ac^l}|W_{n,\Sc}|=F,\quad n \in \Nc.
 \end{align}

It is shown in \cite[Theorem 1]{Jin&Cui:Arxiv2017} that for each file $W_n$, the size of its subfile $W_{n,\Sc}$ only depends on $|\Sc|$.
This implies that $|W_{n,\Sc}|$ is the same for  any $\Sc \in \Ac^l$ of the same size $l$.
Based on this property, for each file $W_n$, its subfiles are grouped  into file subgroups, each denoted  by $\Wc^l_n=\{W_{n,\Sc}: \Sc \in \Ac^l\}$, for $l\in \Kc_0$. There are ${K \choose l}$ subfiles of the same size in  $\Wc^l_n$ (intended for user subsets in cache subgroup $\Ac^l$), and there are total  $K+1$ file subgroups.

Let $a_{n,l}$ denote  the  size of subfiles in $\Wc^l_n$, as a fraction of the file size $F$ bits: $a_{n,l} \triangleq |W_{n,\Sc}|/F$ (for $\forall  \Sc \in \Ac^l$),  $l\in \Kc_0$,   $n\in \Nc$.    Note that $a_{n,0}$ represents the fraction of file $W_n$  that is not stored at any user's cache but only remains in the server. Then, the file partition constraint \eqref{Constraint1} is simplified to
\begin{align}\label{Constraint1.1}
  \sum_{l=0}^{K}{K \choose l}a_{n,l}=1, \quad\; n \in \Nc.
\end{align}

Recall that in file partitioning,  each subfile is intended for a unique user subset. During the cache placement, user $k$  stores all the subfiles in $\Wc_n^l$ that are  intended for user subsets it belongs to, \ie $\{W_{n,\Sc}:  \Sc \in \Ac^l  \text{~and~} k\in \Sc\}\subseteq \Wc_n^l$, for  $l\in \Kc$.
Note that in each $\Ac^l$, $l\in \Kc$, there are total ${K-1 \choose l-1}$  different user subsets containing the same user $k$. Thus,  there are  $\sum_{l=1}^{K}{K-1 \choose l-1}$ subfiles in each file $W_n$ that a user can possibly store in its local cache. With subfile size $a_{n,l}$,  this means  that {each user caches a total of  $\sum_{l=1}^{K}{K-1 \choose l-1}a_{n,l}$ fraction of file $W_n$}. For  cache size $M$ at each user, we have the following local cache  constraint
\begin{align}
\sum_{n=1}^{N}\sum_{l=1}^{K}{K-1 \choose l-1}a_{n,l} \leq M.\label{Constraint2}
\end{align}

We point out that the above construction through subfile and user subset partitioning to represent {an uncoded}  cache placement is general, \ie any {uncoded }cache placement scheme can be equivalently represented by the specific values of $\{a_{n,l}: n\in\Nc,l\in \Kc_0\}$.

\subsubsection{Content Delivery via Coded Multicasting}
  For content delivery by the CCS, the server multicasts  a unique coded message to each user subset. The message is formed by bitwise XOR operation of  subfiles as
\begin{align}\label{coded_msg}
C_\Sc\triangleq \bigoplus_{k \in \Sc} \! W_{d_k,\Sc\backslash\{k\}}.
\end{align}

Note that the  CCS originally proposed in \cite{Maddah-Ali&Niesen:TIT2014} is  shown to be a valid caching scheme for cache size $M=\{0,N/K,2N/K,...,N\}$. This conclusion can be straightforwardly extended to any  cache size  $M$, using the delivery strategy of the decentralized CCS  in \cite{Niesen&Maddah-Ali:TIT2015}. %we show below that the CCS is also a valid coded caching scheme.

With nonuniform file popularities,  the cache placement may be different for files with different popularities. This means the file partitioning may be different among these files, and the subfile size $a_{n,l}$ is a function of $n$. Note that when the sizes of subfiles are not equal, zero padding is needed to code the subfiles together for multicasting in \eqref{coded_msg}. As a result, the size of coded message $C_\Sc$  is determined by the largest subfile among subfiles in the delivery group (user subset) $\Sc$, \ie\begin{align}
   |C_\Sc|=\max_{k\in\Sc}a_{d_k,l}, \ \ \Sc\in\Ac^{l+1}, l=0,\ldots,K-1.\label{CodedMsgPad}
 \end{align}
With  \eqref{coded_msg} and \eqref{CodedMsgPad}, each user in $\Sc$ can retrieve the subfile of its requested file from the coded message $C_\Sc$.

%%%%%%%%%%%%%%%%%%%%%%%%%
\section{Cache Placement Optimization Formulation } \label{sec:problem}

Based on \eqref{CodedMsgPad}, the average rate $\bar R$ of  data delivery  by the CCS is given by
\begin{align}\label{equ:shared_link}
 \bar{R}=\mathbb{E}_\dbf\!\left[\sum_{\Sc \subseteq \Kc, \Sc\neq \emptyset}\!\!\!\!|C_\Sc|\right]
 =\mathbb{E}_\dbf\!\left[\sum_{l=0}^{K-1}\! \sum_{\Sc\in\Ac^{l+1}}\!\max_{k\in\Sc}a_{d_k,l}\right]
\end{align}
where $\mathbb{E}_\dbf[\cdot]$ is taken w.r.t. demand vector $\dbf$.

Let $\abf_n=[a_{n,0},\ldots,a_{n,K}]^T$ denote the $(K+1)\times 1$ cache placement vector for file $W_n, n\in\Nc$.
The cache placement optimization problem  {for the CCS} is formulated  as obtaining  the optimal \{$\abf_n$\} to minimize the average rate $\bar R$, given by\footnote{{Note that {\bf P0} is formulated for the CCS, which is based on uncoded cache placement and one-shot coded delivery with zero padding, as described in Section~\ref{III.A}.}}
\begin{align}
\textrm{\bf P0}: \;\min_{\{\abf_n\}}\;\; & \bar{R} \nn \\
 \textrm{s.t.} \;\; &
\eqref{Constraint1.1},\eqref{Constraint2}, \; \text{and~} \nn\\
&\abf_{n}\succcurlyeq\mathbf{0},\; n\in\Nc. \label{Constraint_gt0}
\end{align}

{The optimization problem {\bf P0} is complicated to solve. In the following, we provide a few simplifications to  the average rate objective and the constraints and transform {\bf P0} into a simplified equivalent problem.}   

\subsection{Problem Reformulation}
For nonuniform file popularities, it is shown that the optimal cache placement under the CCS has a \emph{popularity-first} property \cite{Jin&Cui:Arxiv2017}. Specifically, it states in \cite[Theorem 2]{Jin&Cui:Arxiv2017} that for file popularities $p_1\geq\ldots\geq p_N$, under the optimal cache placement,   the following condition holds for the cached subfiles
  \begin{align}\label{ConstraintPopFir}
a_{n,l}\geq a_{n+1,l},\quad l\in\Kc,\ n\in\Nc\backslash\{N\},
\end{align}
where the amount of cache assigned to a file is monotonic with the file popularity.

Without loss of the optimality,
we  now explicitly impose  constraint \eqref{ConstraintPopFir} and have the following equivalent problem to  {\bf P0}
\begin{align}
\textrm{\bf P1}: \;\min_{\{\abf_n\}}\;\; & \bar R \nn\\
\textrm{s.t.} \;\; & \eqref{Constraint1.1},\eqref{Constraint2},\eqref{Constraint_gt0},  \eqref{ConstraintPopFir}.\nn
\end{align}

At the optimality of {\bf P1}, the local cache constraint \eqref{Constraint2} is attained with equality, \ie the cache memory is always fully utilized. To see this, note that at optimality if there is any unused memory, we can always modify the  assumed optimal caching placement by adding any uncached portion of files into the unused memory. This leads to reduced   $\bar R$, contradicting the assumption that there is unused cache memory at  optimality.
Thus, we replace constraint \eqref{Constraint2} with the  equality constraint
  \begin{align}\label{Constraint1.1_eq}
  \sum_{n=1}^{N}\sum_{l=1}^{K}{K-1 \choose l-1}a_{n,l} = M.
  \end{align}

Next, we show the following lemma for constraint \eqref{Constraint_gt0}.
\begin{lemma}\label{lemma1}
Under constraint \eqref{ConstraintPopFir}, constraint \eqref{Constraint_gt0}  is equivalent to the following two constraints
\begin{align}
&a_{N,l}\geq 0,\  l\in\Kc \label{Constraint_a_Nl}\\
&a_{1,0}\geq 0\label{Constraint_a_10}.
\end{align}
\end{lemma}
\IEEEproof
  If $a_{N,l}\ge 0$,  $\forall l\in\Kc$, by the popularity-first condition \eqref{ConstraintPopFir}, we have
\begin{align}\label{Constraint_a_nl}
a_{n,l}\geq 0, \ \forall l\in\Kc, \ \forall n\in\Nc.
\end{align}
Recall that subfile size  $a_{n,0}$ represents the fraction of  $W_n$ that is not stored at any user cache. From \eqref{Constraint1.1}, we have
\begin{align}\label{ConstraintPopFir_0}
a_{n,0}=1-\sum_{l=1}^{K}{K \choose l}a_{n,l},\ n\in\Nc.
\end{align} Combining  \eqref{ConstraintPopFir} and \eqref{ConstraintPopFir_0}, we have $a_{1,0}\leq\ldots\leq a_{N,0}$.
If $a_{1,0}\geq 0$ in \eqref{Constraint_a_10} holds, then
$a_{n,0}\geq0$, $\forall n\in\Nc$.
 Combining this with \eqref{Constraint_a_nl}, we have $\abf_n\succcurlyeq0$, $\forall n\in \Nc$, which is constraint \eqref{Constraint_gt0}.
\endIEEEproof
By Lemma~\ref{lemma1},   constraints \eqref{Constraint_gt0}  in {\bf P1} can be equivalently replaced by constraints \eqref{Constraint_a_Nl} and \eqref{Constraint_a_10}.

Let $Y_m$, $m=1,\ldots,K$, denote the $m$th smallest file index in the demand vector $\dbf$. The probability distribution of $Y_m$ is obtained in  \cite[Lemma 2]{Daniel&Yu:TIT19} (the expression of $Y_m$ is provided in Appendix \ref{ProbExpression} for completeness).
By the popularity-first property of the optimal cache placement, the average rate $\bar{R}$ in \eqref{equ:shared_link} is shown  to have the following expression~\cite{Daniel&Yu:TIT19} \begin{align}
    \bar{R}=&\sum_{n=1}^{N}\sum_{l=1}^{K-1}\sum_{m=1}^{K}{K-m \choose l}\Pr[Y_m=n]a_{n,l}\nn\\
    &+\sum_{n=1}^{N}\sum_{m=0}^{K-1}\Pr[Y_{K-m}=n]a_{n,0}, \label{equ:AverageRate}
  \end{align}
where $\Pr[Y_m = n]$ is not a function of $\abf_n$.
The above expression shows that $\bar{R}$ is  a weighted sum of $a_{n,l}$'s (for each cache subgroup $l$).

From  \eqref{equ:AverageRate}, define $\gbf_n\triangleq [g_{n,0},\ldots,g_{n,K}]^{\mathrm{T}}$, $n\in \Nc$, where
\begin{align}
g_{n,l}&\triangleq\sum_{m=1}^{K}{K-m \choose l}\Pr[Y_m=n],\quad l\in\Kc, \nn\\
g_{n,0}&\triangleq\sum_{m=0}^{K-1}\Pr[Y_{K-m}=n].\label{ValueC_nl}
\end{align}
Also, from \eqref{Constraint1.1} and \eqref{Constraint1.1_eq}, define $\bbf\triangleq[b_0,\dots,b_K]^T$, with $b_l \triangleq{K \choose l}$, and  $\cbf\triangleq[c_0,\dots,c_K]^T$, with $c_l\triangleq {K-1 \choose l+1}$, $l\in \Kc_0$. Combining the results from \eqref{Constraint1.1_eq} to \eqref{ValueC_nl}, we reformulate the cache placement optimization problem {\bf P1} into the following equivalent LP problem
\begin{align}
\textrm{\bf P2:}\; \min_{\{\abf_n\}}  \;\;
                   &\sum_{n=1}^{N}\gbf_n^T \abf_n\ \nn\\
\textrm{s.t.} \;\;
                   &\eqref{ConstraintPopFir},\eqref{Constraint_a_Nl},\eqref{Constraint_a_10}, \ \text{and}\nn\\
                   &\bbf^T \abf_n=1,\                       n\in\Nc, \label{Constraint_SumTo1}\\
                   &\sum_{n=1}^{N}\cbf^T \abf_n= M\label{Constraint_SumLeM}.
\end{align}

{Note that compared to {\bf P1} with $2N(K+1)-K+1$ constraints, {\bf P2} has  $N(K+1)+2$ constraints. Reducing the constraints facilitates us to explore the Karush-Kuhn-Tucker (KKT) optimality conditions \cite{Boyd2004:ConvexBook} in the problem and obtain the  inherent structure in the optimal cache placement. }

\section{The Optimal Cache Placement}\label{sec:Placement}

In this section, {we derive the optimal cache placement solution to {\bf P2}. We first present a structural property of the optimal cache placement solution for \textrm{\bf P2}. It is obtained by exploring the KKT conditions for {\bf P2}. Based on this property, we identify several possible optimal solution structures. By analyzing each solution structure along with the file partition and cache memory constraints}, we obtain the closed-form  cache placement solution under each solution structure. Finally, we develop a simple low-complexity algorithm using these obtained candidate solutions to obtain the optimal solution for \textrm{\bf P2}.
We first give the definition of \emph{file group} below.
%that is categorized by the cache placement vectors.
%With nonuniform file popularities, files with different popularities could %have different cache placement strategies which are represented by the caching %vectors $\abf_n, n\in\Nc$.

\begin{myDef}
 (\emph{File group}) A file group is a subset of $\Nc$ that contains all files with the same cache placement vector, \ie for any two files $W_n$ and $W_{n^{\prime}}$, if  their placement vectors  $\abf_{n}=\abf_{n^{\prime}}$, then they belong to the same file group.
\end{myDef}

For $N$ files, there could be potential as many as $N$ file groups (\ie all $\abf_n$'s are different), which makes the design of optimal cache placement a major challenge. File grouping is a popular  method proposed for the CCS~\cite{Niesen&Maddah-Ali:TIT2017,Ji&Order:TIT17,Zhang&Coded:TIT18,hachem2017coded} to simplify the cache placement design under nonuniform file popularity. Having fewer file groups reduces the  complexity in determining the placement vectors $\{\abf_n\}$. However,  existing  file grouping schemes are suboptimal.
 Our main result in Theorem~\ref{The3Groups} below describes the structural property, in terms of file groups, of the optimal cache placement  for the CCS.

\begin{theorem}\label{The3Groups}
  For $N$ files with any file popularity distribution  $\pbf$, and for any $K$ and $M\le N$,  there are at most three file groups under the optimal cache placement  $\{\abf_n\}$ for  {\bf P1}.
\end{theorem}

\begin{IEEEproof}
{Since \textbf{P2} is an LP, we explore the KKT conditions for \textbf{P2} to derive the file group property.}   See Appendix \ref{Proof:The3Groups}.
\end{IEEEproof}
\vspace{3pt}

Theorem~\ref{The3Groups} indicates that, regardless of the values of $N$, $\pbf$, $K$, and $M$,   there are only three possible file group structures under the optimal cache placement, \ie one to three file groups. This implies that there are at most three unique vectors among  the optimal cache placement  vectors $\{\abf_n\}$, one for each file group. This property  drastically reduces the complexity in solving the cache placement problem, and in turn, it allows us to explore the solution structure to obtain the optimal  solution $\{\abf_n\}$ analytically.
The result of at most three file groups, regardless of file popularity distribution $\pbf$ among $N$ files, is somewhat surprising. We will provide some insight into this result in Section~\ref{subsubsec:opt_placement}, after the cache placement structure and solution are obtained.

\begin{Remark}
{Existing  file grouping strategies \cite{Niesen&Maddah-Ali:TIT2017,Ji&Order:TIT17,Zhang&Coded:TIT18,hachem2017coded} are either suboptimal or designed for a specific file popularity distribution.} {Some of these  suboptimal file grouping strategies \cite{Ji&Order:TIT17,Zhang&Coded:TIT18,hachem2017coded} were shown to be a constant factor away from the optimum  in terms of the average rate. Since the constant factor is relatively large, it remains unclear how close their performance is to that under the optimal cache placement strategy for the CCS.} Furthermore, under a file grouping strategy, the specific cache placement for each group is needed. Existing works use the symmetric decentralized cache placement strategy for each group. In contrast, by Theorem \ref{The3Groups}, in the following, we will discuss each of the three file grouping cases to obtain the corresponding optimal placement.
\end{Remark}

Following Theorem~\ref{The3Groups}, we will examine all three cases of file groups  for   {\bf P2} to obtain the placement solution. We first introduce the following notations to be used later:
\begin{itemize}
\item Denote $\bar{\abf}_{n}=[a_{n,1},\ldots,a_{n,K}]^T$as  the sub-placement vector in $\abf_{n}$. It specifies only the size of each subfile stored in the local cache, while $a_{n,0}$ specifies the subfile kept at the server.
\item We use notation  $\bar\abf_n\succcurlyeq_1 {\bf 0}$ to indicate that there is at least one positive element in $\bar\abf_n$; otherwise,  $\bar\abf_n= {\bf 0}$.
Similarly,  $\bar{\abf}_{n_1}\succcurlyeq_1\bar{\abf}_{n_2}$ denotes that  at least one element in $\bar\abf_{n_1}$ is greater than that in $\bar\abf_{n_2}$, and all the rest elements in  $\bar\abf_{n_1}$ and  $\bar\abf_{n_2}$ are equal.
\end{itemize}

With the above notations, we establish the following  equivalence on the placement vectors:
\begin{enumerate}
\item By \eqref{Constraint1.1}, for any two files $n_1$ and $n_2$, we have
\begin{align}\label{equ:PlacementToSub1}
\abf_{n_1}=\abf_{n_2} \ \Leftrightarrow \ \bar\abf_{n_1}=\bar\abf_{n_2}
\end{align}
where ``$\Leftrightarrow$'' denotes being equivalent.
\item By \eqref{ConstraintPopFir} and \eqref{ConstraintPopFir_0}, for any two files with their indexes $n_1<n_2$, we have
\begin{align}\label{equ:PlacementToSub}
  \abf_{n_1}\neq\abf_{n_2} \ \Leftrightarrow \ \bar\abf_{n_1}\succcurlyeq_1\bar\abf_{n_2}\text{~and~} a_{n_1,0}<a_{n_2,0}.
\end{align}
\end{enumerate}

In the following, we consider each case of file groups,  and  identify the complete structure of the cache placement vector and obtain the optimal solution for this case.

%%%%%%%%%%%%%%%%%%%%%%%%%%%%%%%%%%%%
\subsection{One File Group}\label{sec:oneGroup}

With a single file group, the cache placement vectors are the same for all files.
Let $\abf_1=\cdots=\abf_N=\abf$. In this case, we can simplify the expressions in \textrm{\bf P2}. Denote $\tilde\gbf\triangleq[\tilde{g}_0,\ldots,\tilde{g}_K]^T$ with $\tilde g_l=\binom{K}{l+1}, l\in \Kc_0$. Then, \textrm{\bf P2} is simplified into the following equivalent problem
\begin{align}
\text{\bf P3:}\; \min_{\abf}  \;\; &\tilde\gbf^T \abf\ \nn\\
\text{s.t.} \;\;
                   &\bbf^T \abf=1, \label{SimpEqualCon:1}\\
                   &\cbf^T \abf= \frac{M}{N}, \label{SimpUnequalCon:1}\\
                   &\abf\succcurlyeq \mathbf{0}\label{SimpUnequalCon:2}.
\end{align}

Note that {\bf P3} is the same as  the cache placement optimization problem for the  uniform file popularity case (the same placement vector $\abf$ for all files), of which the optimal solution has been obtained in \cite{Daniel&Yu:TIT19} in closed-form. To summarize, the optimal  $\abf$  for {\bf P3} is given as follows:\begin{enumerate}
\item[i)] If $MK/N \in \mathbb{N}$: The optimal  $\abf$ has only one nonzero element: $a_{l_o}=1/{K \choose l_o}$,  $l_o =MK/N $, and $a_l=0$,  $\forall \ l\neq l_o$.
\item[ii)] If $MK/N \notin \mathbb{N}$: The optimal  $\abf$ has two nonzero adjacent elements: Let $v\triangleq \frac{KM}{N}$. Then,
\begin{align}\label{one_group:a_opt}
&a_{l_o}=\frac{1+\left\lfloor v\right\rfloor- v}{{K \choose \left\lfloor v\right\rfloor}}, \ \
 a_{l_o+1}=\frac{v-\left\lfloor v\right\rfloor}{{K \choose \left\lceil v\right\rceil}}, \quad l_o=\left\lfloor v\right\rfloor \nn \\
& a_l=0, \quad \forall \ l \neq l_o \ \text{or} \ l_o+1.
\end{align}
\end{enumerate}
Note that Case i) is  a special case of Case ii): In Case ii),  if $l_o =  v$, $a_{l_o+1}=0$, the solution  in  \eqref{one_group:a_opt} reduces to that of case i). Thus, the optimal solution of {\bf P3} can be simply summarized in \eqref{one_group:a_opt}.

\begin{figure}[t]
 \psfrag{Subgroups of a file}[][][0.9]{\quad File subgroups}
  \psfrag{1st File Group}[][][0.75]{1st File Group}
  \psfrag{2nd File Group}[][][0.75]{2nd File Group}
  \psfrag{File 1}[][][0.8]{File 1}
  \psfrag{$N$}[][][0.8]{$N$}
    \psfrag{One File Group}[][][0.85]{One File Group}
  \psfrag{$a_{1,0}$}[][][0.8]{$a_{1,0}$}
  \psfrag{$a_{1,l_o}$}[][][0.8]{$a_{1,l_o}$}
  \psfrag{$a_{1,l_o+1}$}[][][0.8]{$a_{1,l_o+1}$}
  \psfrag{$a_{1,K}$}[][][0.8]{$a_{1,K}$}
    \psfrag{$a_{N,0}$}[][][0.8]{$a_{N,0}$}
  \psfrag{$a_{N,l_o}$}[][][0.8]{$a_{N,l_o}$}
  \psfrag{$a_{N,l_o+1}$}[][][0.7]{$a_{N,l_o+1}$}
  \psfrag{$a_{N,K}$}[][][0.8]{$a_{N,K}$}
  \centering
  \includegraphics[scale=0.35]{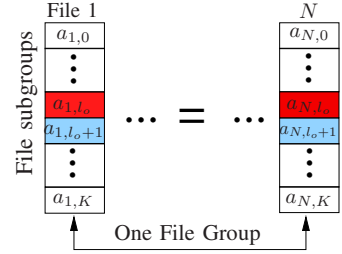}
  \caption{An example of the optimal cache placement for one file group:  $\abf_n=\abf$, $\forall n$,  with $a_{l_o},a_{l_o+1}>0$ and $a_l=0$,  $\forall l\neq l_o,l_o+1$. (The same color indicates the same value of $a_l$)}
  \label{fig:OneGroup}
\end{figure}

The above shows that the optimal $\abf$ has  at most two nonzero elements.  When $MK/N$ is an integer,  $\abf$ has only one nonzero element, which means each file is partitioned into equal subfiles of size $a_{l_o}$. Otherwise,  $\abf$ has two nonzero adjacent elements, which means each file is partitioned into subfiles of two different sizes $a_{l_o}$ and $a_{l_o+1}$. Each subfile is cached into its intended user subset of size $l_o$ or $l_o+1$, as described in Section~\ref{III.A}.  Fig.~\ref{fig:OneGroup} illustrates the optimal $\abf$ in the one-file-group case.

\subsection{Two File Groups}\label{sec:TwoGroups}

For the case of two file groups, there are only two unique placement vectors in  $\{\abf_n\}$. By \eqref{ConstraintPopFir}, this implies that $\{\abf_n\}$ has the following structure:
$\abf_1=\ldots=\abf_{n_o} \neq \abf_{n_o+1}=\ldots=\abf_{N}$, for some $n_o\in \{1,\ldots, N-1\}$.
By \eqref{equ:PlacementToSub1} and \eqref{equ:PlacementToSub}, this  is equivalent to\begin{align}\label{2group_a}
\begin{cases}\bar\abf_1=\ldots=\bar\abf_{n_o} \succcurlyeq_1 \bar\abf_{n_o+1}=\ldots=\bar\abf_{N} \\
a_{1,0}=\ldots=a_{n_o,0}<a_{n_o+1,0}=\dots=a_{N,0}
\end{cases}
\end{align}
for some $n_o\in\{1,\ldots,N-1\}$.
It immediately follows that $a_{n_o+1,0}= \dots=a_{N,0}>0$.
We use $\abf_{n_o}$ and $\abf_{n_o+1}$ to represent the two  unique placement vectors for the first and the second file group, respectively.
We first characterize the structure of the placement vector $\abf_{n_o+1}$ for the second file group below.
\begin{proposition}\label{Pro:TwoGroupOneDiffEle1}
  If there are two file groups under the optimal cache placement $\{\abf_n\}$, the optimal sub-placement vector $\bar\abf_{n_o+1}$ for the second file group has at most one nonzero element.
\end{proposition}
\IEEEproof
See Appendix \ref{ProofPro:TwoGroupOneDiffEle1}.
\endIEEEproof

 Proposition \ref{Pro:TwoGroupOneDiffEle1} indicates that either  $\bar\abf_{n_o+1}=\mathbf{0}$  or $\bar{\abf}_{n_o+1}$ has only one nonzero element. For the former, it means the files in the second file group are not cached but remain at the server only. Note that   two file groups  were considered for placement strategies in \cite{Ji&Order:TIT17,Zhang&Coded:TIT18}, where  the second file group containing less popular files  remains at the server, and the location of $n_o$ for the grouping was
proposed in different heuristic ways. These file grouping methods fall into the case of  $\bar\abf_{n_o+1}=\mathbf{0}$.  However,   the case of allocating cache to the second file group, \ie $\bar\abf_{n_o+1}\neq\mathbf{0}$,  has never been considered in the  literature.

Following Proposition \ref{Pro:TwoGroupOneDiffEle1}, we  obtain the optimal cache placement in each of the two cases for $\bar{\abf}_{n_o+1}$ below:

\subsubsection{$\bar{\abf}_{n_o+1}=\mathbf{0}$}\label{subsec:ZeroFirst}
\begin{figure}
  \psfrag{Subgroups of a file}[][][0.9]{\quad File subgroups}

  \psfrag{File 1}[][][0.8]{File $1$}
  \psfrag{$n_o$}[][][0.8]{$n_o$}
  \psfrag{$n_o+1$}[][][0.8]{$n_o+1$}
  \psfrag{$N$}[][][0.8]{$N$}

  \psfrag{1st File Group}[][][0.75]{1st File Group}
  \psfrag{2nd File Group}[][][0.75]{2nd File Group}
  \psfrag{$a_{1,0}$}[][][0.8]{$a_{1,0}$}
  \psfrag{$a_{1,l_o}$}[][][0.8]{$a_{1,l_o}$}
  \psfrag{$a_{1,l_o+1}$}[][][0.8]{$a_{1,l_o+1}$}
  \psfrag{$a_{1,K}$}[][][0.8]{$a_{1,K}$}

    \psfrag{$a_{n_o,0}$}[][][0.8]{$a_{n_o,0}$}
  \psfrag{$a_{n_o,l_o}$}[][][0.8]{$\ a_{n_o,l_o}$}
  \psfrag{$a_{n_o,l_o+1}$}[][][0.8]{$\ a_{n_o,l_o+1}$}
  \psfrag{$a_{n_o,K}$}[][][0.8]{$a_{n_o,K}$}

      \psfrag{$a_{n_o+1,0}$}[][][0.75]{$a_{n_o+1,0}$}
  \psfrag{$a_{n_o+1,l_o}$}[][][0.75]{$a_{n_o+1,l_o}$}
  \psfrag{$a_{n_o+1,l_o+1}$}[][][0.6]{$\ a_{n_o+1,l_o+1}$}
  \psfrag{$a_{n_o+1,K}$}[][][0.75]{$a_{n_o+1,K}$}

    \psfrag{$a_{N,0}$}[][][0.8]{$a_{N,0}$}
  \psfrag{$a_{N,l_o}$}[][][0.8]{$a_{N,l_o}$}
  \psfrag{$a_{N,l_o+1}$}[][][0.8]{$a_{N,l_o+1}$}
  \psfrag{$a_{N,K}$}[][][0.8]{$a_{N,K}$}
\centering
  \includegraphics[scale=0.4]{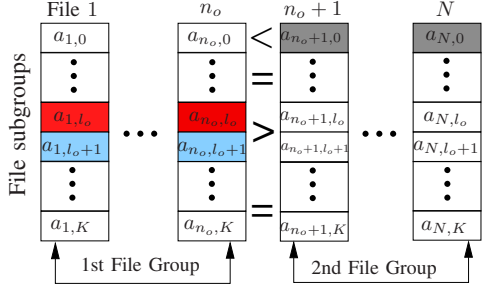}
  \caption{An example of the optimal cache placement for two file groups with $\bar{\abf}_{n_o+1}={\bf 0}$. The 1st file group: $a_{n,l_o}>0$, $a_{n,l_o+1}>0$, for $n=1,\ldots,n_o$, and the rest are all $0$'s. The second file group: $a_{n_o+1,0}=\cdots=a_{N,0}=1$.}
  \label{fig:TwoGroup1}
\end{figure}

\begin{figure}[t]
 % \psfrag{Rate}{$\bar R_{\textrm{uni}}$}
  \centering
  \includegraphics[scale=0.4]{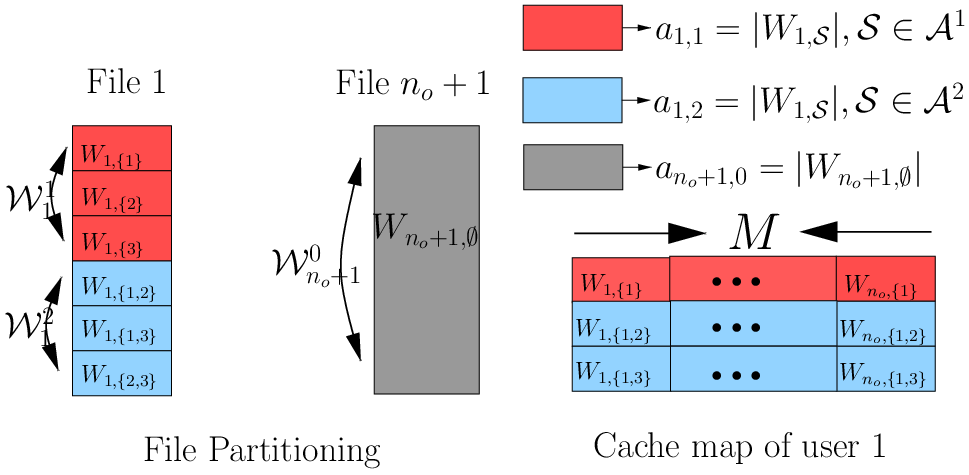}
  \caption{An illustration of file partition and cache placement based on the placement structure  in Fig.~\ref{fig:TwoGroup1}, for $K=3$ users, and $l_o=1$. File $W_1$ in the 1st file group is partitioned into subfiles of two sizes $a_{1,1}$ and $a_{1,2}$. Subfiles in file subgroup   $\Wc_{1}^1$  with  size $a_{1,1}=|W_{1,\Sc}|/F$ (red) is placed in user subset $\Sc\in\Ac^1=\{\{1\},\{2\},\{3\}\}$; Subfiles  in file subgroup  $\Wc_{1}^2$   with size $a_{1,2}=|W_{1,\Sc}|/F$ (blue) is placed in user subset $\Sc\in\Ac^2=\{\{1,2\},\{1,3\},\{2,3\} \}$. For file $W_{n_o+1}$ in the second file group, the entire file is stored solely in the server: $W_{n_o+1,\emptyset}=W_{n_o+1}$,~$a_{n_o+1,0}=1$. The cache memory map of user $1$ shows the stored subfiles of the 1st file group $\{W_1,\ldots,W_{n_o}\}$.} \label{fig:plcmt_demo}
\end{figure}

By \eqref{Constraint1.1}, we have  $a_{n_o+1,0}=1$.
It means that  no cache is allocated to the second file group, and the entire cache  is given to the first file group. It follows that the cache placement problem for $\abf_{n_o}$ of the first group is reduced to that in the one-file-group case in Section~\ref{sec:oneGroup}.
Specifically, %let $\abf_{1}=\ldots=\abf_{n_o}=\abf$.
we can  treat the first file group as a new database consisting of these $n_o$  files, for some $n_o\in\{1,\ldots,N-1\}$. Then, the cache placement optimization problem for $\abf_{n_o}$ is the same as {\bf P3}, except that $N$ is replaced by $n_o$ in constraint \eqref{SimpUnequalCon:1}.
It follows that, the optimal solution is the same as in \eqref{one_group:a_opt}, except that   $N$ is replaced by $n_o$, and $v = MK/n_o$, \ie
\begin{align}\label{two_group1:a_opt}
\hspace*{-1.5em}\begin{cases}\displaystyle a_{n_o,l_o}=\frac{1+\left\lfloor v\right\rfloor- v}{{K \choose \left\lfloor v\right\rfloor}}, \
 a_{n_o,l_o+1}=\frac{v-\left\lfloor v\right\rfloor}{{K \choose \left\lceil v\right\rceil}}, \ l_o=\left\lfloor v\right\rfloor \\
 a_{n_o,l}=0, \quad \forall \ l \neq l_o \ \text{or} \ l_o+1.
 \end{cases} \hspace*{-2em}
\end{align}
An example of the placement $\{\abf_n\}$ of files in this case is shown in Fig.~\ref{fig:TwoGroup1}, where  $\abf_{n_o}$ for the first file group has two adjacent nonzero elements.  In addition,  for this case,  Fig.~\ref{fig:plcmt_demo} illustrates the actual file partitions and cached contents in  user 1.


\begin{algorithm}[t]
\caption{The Cache Placement for the Extended Two-File-Group Case with $\bar\abf_{n_o+1}=\mathbf{0}$ (including one file group)}\label{alg:twogroup1}
\begin{algorithmic}[1]
\Require
   { $K$,  $M$,  $N$, and  $\pbf$.}
\Ensure
    {($\bar R_{\min}$, $n_o^{*}$)}
%\State Set $\gbf$, $\bbf$, $\cbf$;
\For{$n_o$ = $1$ to $N$}
    \State Set $l_o=\lfloor\frac{MK}{n_o}\rfloor$; Set $\bar{\abf}_{n_o+1}=\mathbf{0}$, if $n_o<N$.
    \State Determine  $\abf_{n_o}$ by \eqref{two_group1:a_opt}.
\State Compute $\bar{R}_1(n_o)$ using \eqref{equ:AverageRate}, by replacing $N$ with $n_o$ in \eqref{equ:AverageRate}.
\EndFor
\State Compute $\bar n_o^{*}=\textrm{argmin}_{n_o\in\Nc}\bar R_1(n_o)$; Set $\bar{R}_{\min}=\bar{R}_{1}(n_o^*)$.
\end{algorithmic}
\end{algorithm}

Based on the similarity of the solutions in \eqref{one_group:a_opt} and \eqref{two_group1:a_opt}, we  can extend the two-file-group case to also include  one file group as a special case where $n_o=N$. As a result, for the extended two-file-group case, the optimal cache placement solution is given by \eqref{two_group1:a_opt}, for $n_o\in \{1,\ldots,N\}=\Nc$.
What remains is to obtain the optimal $n_o^*$ to determine $\{\abf_n\}$ that minimizes the average rate objective in {\bf P2}. The optimal $n_o^*$ is the location to determine the file groups. It depends on $(N,\pbf, M,K)$ and  is challenging to obtain analytically. Nonetheless, $\bar{R}$ can be easily computed using \eqref{two_group1:a_opt} for $n_o\in\Nc$, and we can conduct a search for $n_o$ to  determine $n_o^*$ that gives the minimum $\bar{R}$. The algorithm to obtain the placement solution $\{\abf_n\}$ in this case is summarized in Algorithm \ref{alg:twogroup1}.
Through a 1-D search for the optimal $n_o^*$, the algorithm computes $\bar{R}$ using the closed-form expression in \eqref{equ:AverageRate} by $N$ times.

%%%%%%%%%%%%%
\subsubsection{$\bar\abf_{n_o+1}\succcurlyeq_1 \mathbf{0}$}\label{subsec:oneZeroSecond} In this case, by Proposition \ref{Pro:TwoGroupOneDiffEle1},  $\bar{\abf}_{n_o+1}$  has only one nonzero element. Assume $a_{n_o+1,l_o}>0$, for some $l_o\in \Kc$, and $a_{n_o+1,l}=0$, $\forall l\neq l_o$, $l\in \Kc$.
We have the following propositions describing the properties of $\abf_{n_o}$ and $\abf_{n_o+1}$. Proposition~\ref{Pro:TwoGroupsTwoUnequals} specifies the differences of $\bar{\abf}_{n_o}$ and $\bar{\abf}_{n_o+1}$ for the two file groups, and Proposition~\ref{Pro:TwoGroupsFirstZero} characterizes the placement $\abf_{n_o}$ for the first file group.

\begin{proposition}\label{Pro:TwoGroupsTwoUnequals}
  If there are two file groups under the optimal cache placement $\{\abf_n\}$, and $\bar\abf_{n_o+1}\succcurlyeq_1\mathbf{0}$, for some $n_o\in \{1,\ldots,N-1\}$, then  $\bar\abf_{n_o}$ and $\bar\abf_{n_o+1}$ are different by only one element.
\end{proposition}
\IEEEproof
See Appendix \ref{ProofPro:TwoGroupsTwoUnequals}.\endIEEEproof

\begin{proposition}\label{Pro:TwoGroupsFirstZero}
  If there are two file groups under the optimal cache placement $\{\abf_n\}$, and $\bar\abf_{n_o+1}\succcurlyeq_1\mathbf{0}$, for some $n_o\in \{1,\ldots,N-1\}$, then $a_{n_o,0}=0$.
\end{proposition}
\IEEEproof
See Appendix \ref{ProofPro:TwoGroupsFirstZero}.
\endIEEEproof

\begin{figure}[t]
  \psfrag{Subgroups of a file}[][]{File subgroups}
  \psfrag{1st File Group}[][][0.8]{1st File Group}
  \psfrag{2nd File Group}[][][0.8]{2nd File Group}

  \psfrag{File 1}[][][0.8]{File $1$}
  \psfrag{$n_o$}[][][0.8]{$n_o$}
  \psfrag{$n_o+1$}[][][0.8]{$n_o+1$}
  \psfrag{$N$}[][][0.8]{$N$}

  \psfrag{1st File Group}[][][0.75]{1st File Group}
  \psfrag{2nd File Group}[][][0.75]{2nd File Group}
  \psfrag{$a_{1,0}$}[][][0.8]{$a_{1,0}$}
  \psfrag{$a_{1,l_o}$}[][][0.8]{$a_{1,l_o}$}
  \psfrag{$a_{1,K}$}[][][0.8]{$a_{1,K}$}

    \psfrag{$a_{n_o,0}$}[][][0.8]{$a_{n_o,0}$}
  \psfrag{$a_{n_o,l_o}$}[][][0.8]{$\ a_{n_o,l_o}$}
  \psfrag{$a_{n_o,K}$}[][][0.8]{$a_{n_o,K}$}

      \psfrag{$a_{n_o+1,0}$}[][][0.75]{$a_{n_o+1,0}$}
  \psfrag{$a_{n_o+1,l_o}$}[][][0.75]{$a_{n_o+1,l_o}$}
  \psfrag{$a_{n_o+1,K}$}[][][0.75]{$a_{n_o+1,K}$}

    \psfrag{$a_{N,0}$}[][][0.8]{$a_{N,0}$}
  \psfrag{$a_{N,l_o}$}[][][0.8]{$a_{N,l_o}$}
  \psfrag{$a_{N,K}$}[][][0.8]{$a_{N,K}$}
\centering
  \includegraphics[scale=0.4]{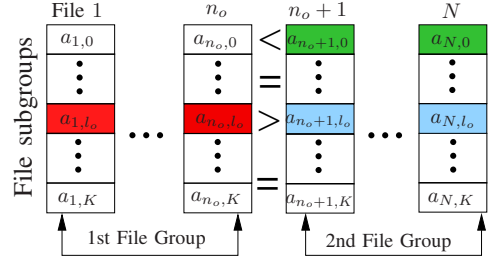}
\caption{An example of the optimal cache placement for two file groups with $\bar{\abf}_{n_o+1}\succcurlyeq_1 0$: i) $0=a_{n_o,0}<a_{n_o+1,0}<1$. ii) Between $\bar\abf_{n_o}$ and $\bar\abf_{n_o+1}$: $a_{n_o,l_o}>a_{n_o+1,l_o}>0$; $a_{n_o,l}=a_{n_o+1,l}=0$, $\forall l\in \Kc, l\neq l_o$.}
  \label{fig:TwoGroup2}
\end{figure}

\begin{figure}[t]
  \psfrag{Subgroups of a file}[][]{\quad File subgroups}
  \psfrag{1st File Group}[][][0.8]{1st File Group}
  \psfrag{2nd File Group}[][][0.8]{2nd File Group}

  \psfrag{File 1}[][][0.8]{File $1$}
  \psfrag{$n_o$}[][][0.8]{$n_o$}
  \psfrag{$n_o+1$}[][][0.8]{$n_o+1$}
  \psfrag{$N$}[][][0.8]{$N$}

  \psfrag{$a_{1,0}$}[][][0.8]{$a_{1,0}$}
  \psfrag{$a_{1,l_o}$}[][][0.8]{$a_{1,l_o}$}
  \psfrag{$a_{1,l_1}$}[][][0.8]{$a_{1,l_1}$}
  \psfrag{$a_{1,K}$}[][][0.8]{$a_{1,K}$}

  \psfrag{$a_{n_o,0}$}[][][0.8]{$a_{n_o,0}$}
  \psfrag{$a_{n_o,l_o}$}[][][0.8]{$\ a_{n_o,l_o}$}
  \psfrag{$a_{n_o,l_1}$}[][][0.8]{$a_{n_o,l_1}$}
  \psfrag{$a_{n_o,K}$}[][][0.8]{$a_{n_o,K}$}

  \psfrag{$a_{n_o+1,0}$}[][][0.8]{$a_{n_o+1,0}$}
  \psfrag{$a_{n_o+1,l_o}$}[][][0.76]{$\ a_{n_o+1,l_o}$}
  \psfrag{$a_{n_o+1,l_1}$}[][][0.76]{$\ a_{n_o+1,l_1}$}
  \psfrag{$a_{n_o+1,K}$}[][][0.76]{$a_{n_o+1,K}$}

    \psfrag{$a_{N,0}$}[][][0.8]{$a_{N,0}$}
  \psfrag{$a_{N,l_o}$}[][][0.8]{$a_{N,l_o}$}
  \psfrag{$a_{N,l_1}$}[][][0.8]{$a_{N,l_1}$}
  \psfrag{$a_{N,K}$}[][][0.8]{$a_{N,K}$}  
  
  \centering
  \includegraphics[scale=0.4]{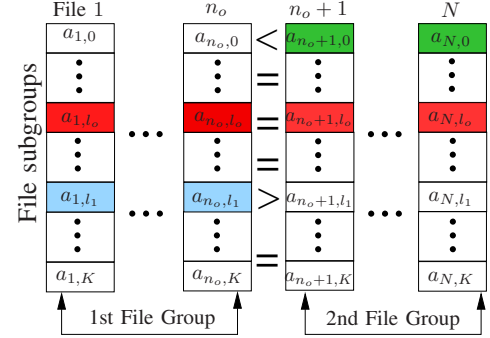}
  \caption{An example of the optimal cache placement $\{\abf_n\}$ in the case of two file groups with $\bar{\abf}_{n_o+1}\succcurlyeq_1 \mathbf{0}$: $a_{n_o+1,0}>a_{n_o,0}=0$. Between $\bar\abf_{n_o}$ and $\bar\abf_{n_o+1}$: 1) $a_{n_o,l_1}>a_{n_o+1, l_1}=0$; 2) $a_{n_o,l_o}=a_{n_o+1,l_o}>0$; 3) $a_{n_o,l}=a_{n_o+1, l}=0$, $\forall l\in \Kc$,~$l\neq l_o,l_1$.}
  \label{fig:TwoGroup3}
\end{figure}

Proposition~\ref{Pro:TwoGroupsFirstZero} indicates that each file in the  first file group has all its subfiles cached  among  $K$ users, and no  subfile  solely remains in the server.
Recall in this case that  $\bar{\abf}_{n_o+1}$ has only one nonzero element $a_{n_o+1,l_o}>0$. By Proposition \ref{Pro:TwoGroupsTwoUnequals},
the different element between $\bar\abf_{n_o}$ and $\bar\abf_{n_o+1}$ can be either at index  $l_o$ or some $l_1$, for $l_1\neq l_o$. By the popularity-first property in \eqref{ConstraintPopFir}, either  of the following two cases holds: {2.i)} $a_{n_o,l_o}>a_{n_o+1,l_o}>0$; or {2.ii)} $a_{n_o,l_1}>a_{n_o+1,l_1}=0$, for some   $l_1\neq l_o$, $l_1\in\Kc$.
The structure of $\{\abf_n\}$ in Case 2.i) and Case 2.ii) is illustrated in Figs.~\ref{fig:TwoGroup2} and~\ref{fig:TwoGroup3}, respectively. We point out that $l_o$ and $l_1$ are not necessarily adjacent to each other.
Now we derive the solution $(\abf_{n_o},\abf_{n_o+1})$ in  each of these two  cases:

\vspace*{0.5em}
\noindent \emph{Case 2.i)}  $a_{n_o,l_o}>a_{n_o+1,l_o}>0$:\\
In this case,  $\bar\abf_{n_o}$ and $\bar\abf_{n_o+1}$ are only different at the $l_o$th nonzero element in $\bar\abf_{n_o+1}$. It follows that  $a_{n_o,l}=a_{n_o+1,l}=0$, $\forall l\neq l_o$, $l\in\Kc$. By Proposition \ref{Pro:TwoGroupsFirstZero}, we conclude that $a_{n_o,l_o}$ is the only nonzero element in $\abf_{n_o}$.
From \eqref{Constraint_SumTo1} and \eqref{Constraint_SumLeM}, we have
\begin{align}\label{equ:twogroupscon1}
b_{l_o}a_{n_o,l_o}&=1 \nn\\
     n_oc_{l_o}a_{n_o,l_o}+(N-n_o)c_{l_o}a_{n_o+1,l_o}&=M.
\end{align}
Solving  \eqref{equ:twogroupscon1} and substituting the expressions of    $b_{l_o}$ and $c_{l_o}$ defined below \eqref{ValueC_nl}, we have % determine $\abf_{n_o}$ and $\abf_{n_o+1}$
\begin{align}\label{equ:Twogroups2.i}
\!a_{n_o,l_o}&\!=\frac{1}{{K \choose l_o}}, \ \  %\label{equ:Twogroups2}\\
 a_{n_o+1,l_o}=\frac{1}{{K \choose l_o}}\left(\!\frac{\frac{KM}{l_oN}-\frac{n_o}{N}}{1-\frac{n_o}{N}}\!\right).
\end{align}
By the condition of Case 2.i)  $a_{n_o,l_o}>a_{n_o+1,l_o}>0$, \eqref{equ:Twogroups2.i} is only valid if $n_o < KM/l_o<N$, for $l_o \in \Kc$. Thus, the range of $l_o$ for this case to be a valid candidate for  the optimal placement is
\begin{align}\label{case2i:l_o}
\left\lfloor\frac{KM}{N}\right\rfloor+1 \le  l_o \le \min \left\{K, \left\lceil\frac{KM}{n_o}\right\rceil-1\right\}.
\end{align}
Finally, $a_{n,0}$'s can be obtained by \eqref{ConstraintPopFir_0}. To summarize, the placement solution $(\abf_{n_o},\abf_{n_o+1})$  in this case is given by
\begin{align}
&a_{n_o,l_o}=\frac{1}{{K \choose l_o}}, \quad a_{n_o,l}=0, \ \forall \ l\neq l_o \label{equ:Twogroups2i_1}\\
&\hspace*{-.7em}\begin{cases}\displaystyle a_{n_o+1,0}\!=\frac{1-\frac{KM}{l_oN}}{1-\frac{n_o}{N}}, \  \
a_{n_o+1,l_o} \!= \frac{1}{{K \choose l_o}}\!\!\left(\!\frac{\frac{KM}{l_oN}-\frac{n_o}{N}}{1-\frac{n_o}{N}}\!\!\right)
   \\
a_{n_o+1,l}= 0,  \ \forall \ l\neq 0 \ \text{or} \ l_o
\end{cases}  \label{equ:Twogroups2i_2}
\end{align}
where $l_o$ satisfies \eqref{case2i:l_o}, and $n_o \in \{1,\ldots, N-1\}$.

Fig.~\ref{fig:TwoGroup2} illustrates the above result in this case under  two file groups as the optimal placement, where different color blocks indicate the  different values of $\{a_{n,l}\}$.

\vspace*{0.5em}
\noindent \emph{Case 2.ii)} $a_{n_o,l_1}>a_{n_o+1,l_1}=0$,  $l_1\neq l_o$:\\ In this case, the $l_o$th element in     $\bar\abf_{n_o}$ and $\bar\abf_{n_o+1}$ are identical, and we have $a_{n_o,l_o}=a_{n_o+1,l_o}>0$. Since $a_{n_o,0}=0$ by Proposition \ref{Pro:TwoGroupsFirstZero}, we conclude that
   $\abf_{n_o}$ has two nonzero elements $a_{n_o,l_o}$ and $a_{n_o,l_1}$.  Also, recall from \eqref{2group_a} that $a_{n_o+1,0}>0$. Thus, $\abf_{n_o+1}$ has two nonzero elements  $a_{n_o+1,0}$ and $a_{n_o+1,l_o}$. The rest elements $\abf_{n_o}$ and $\abf_{n_o+1}$ are all zeros.
The placement structure of $\{\abf_n\}$ in this case  is illustrated in Fig.~\ref{fig:TwoGroup3}, where nonzero elements in $\abf_{n_o}$ and $\abf_{n_o+1}$ are shown as colored blocks and zero elements as uncolored blocks. Given the structure of $\abf_{n_o}$ and $\abf_{n_o+1}$, by \eqref{Constraint_SumTo1} and \eqref{Constraint_SumLeM}, we have
      \begin{align}
&b_{l_o}a_{n_o,l_o}+b_{l_1}a_{n_o,l_1}=1, \ %\label{equ:twogroupscon2.1}
a_{n_o+1,0}+b_{l_o}a_{n_{o},l_o}=1 \label{equ:twogroupscon2.2} \\
&Nc_{l_o}a_{n_o,l_o}+n_oc_{l_1}a_{n_o,l_1}=M.\label{equ:twogroupscon2.3}
\end{align}
Solving \eqref{equ:twogroupscon2.2} and \eqref{equ:twogroupscon2.3}, and substituting the expressions of    $b_l$ and $c_l$ given below \eqref{ValueC_nl}, we obtain the solution of $(\abf_{n_o},\abf_{n_o+1})$ as
\begin{align}
&\hspace*{-1em}\begin{cases}\displaystyle a_{n_o,l_o}=\frac{1}{{K \choose l_o}}\frac{\frac{KM}{l_oN}-\frac{l_1n_o}{l_oN}}{1-\frac{l_1n_o}{l_oN}},
\
a_{n_o,l_1}=\frac{1}{{K \choose l_1}}\frac{1-\frac{KM}{l_oN}}{1-\frac{l_1n_o}{l_oN}}
\\
a_{n_o,l}=0, \ \forall \ l\neq l_o \ \text{or} \ l_1,
\end{cases} \!\!\!\!\!\!\label{equ:Twogroup2ii_1} \\
&\hspace*{-1em}\begin{cases}\displaystyle a_{n_o+1,l_o}=a_{n_o,l_o}, \ \  \quad  a_{n_o+1,0}=\frac{1-\frac{KM}{l_oN}}{1-\frac{l_1n_o}{l_oN}}  \\
a_{n_o+1,l}= 0, \forall \ l\neq 0 \ \text{or} \ l_o  \end{cases}\label{equ:Twogroup2ii_2}
\end{align}
where for $a_{n_o,l_o}$, $a_{n_o,l_1}$, and $a_{n_o+1,0}$ being all positive,  $l_o$ and $l_1$ should satisfy one of the following constraints
\begin{itemize}
\item[C1)] $l_o > KM/N$ and $ l_1 < KM/n_o$, or
\item[C2)] $ l_o<KM/N$ and $l_1 > KM/n_o$.
\end{itemize}
Note that, if $n_o\le M$, only constraint (C1) is valid.


In summary, for the case of two file group with $\bar{\abf}_{n_o+1}\succcurlyeq_1 \mathbf{0}$, by \eqref{2group_a}, the placement   $\{\abf_n\}$ are determined via  $(\abf_{n_o},\abf_{n_o+1})$ in Cases 2.i) and 2.ii) for given $(n_o,l_o)$ or $(n_o,l_o,l_1)$, respectively. Since $(n_o,l_o)$ can be viewed as a special case of $(n_o,l_o,l_1)$ for $l_1=l_o$, to unify the notations for different cases, we define $(n_o,l_o,l_o) \triangleq (n_o,l_o)$. As a result, the average rate $\bar{R}$ in {\bf P2} is  a function of $(n_o,l_o,l_1)$. To obtain the best tuple $(n_o,l_o,l_1)$  that results in minimum  $\bar{R}$, we can  search over all possible values of $n_o\in\{1,\ldots,N-1\}$ and $l_o,l_1\in \Kc$ within their respective range constraint in each case.
The detail of obtaining the best solution $\{\abf_n\}$ is summarized in Algorithm \ref{alg:twogroup2}. In the algorithm, we express $\bar{R}$ explicitly as $\bar{R}(n_o,l_o,l_1)$ to emphasize its dependency on $(n_o,l_o,l_1)$. It computes $\bar{R}(n_o,l_o,l_1)$ using the closed-form expression in  \eqref{equ:AverageRate}  for at most   $(N-1)K^2$ times in the worst case at different $(n_o,l_o,l_1)$, which can be done in parallel. Thus, the complexity of the algorithm is very low.
\begin{algorithm}[t]
\caption{The Cache Placement for Two File Groups with $\bar\abf_{n_o+1}\succcurlyeq_1 \mathbf{0}$}\label{alg:twogroup2}
\begin{algorithmic}[1]
\Require
  { $K$,  $M$,  $N$, and  $\pbf$}
\Ensure
    {($\bar R_{\min},n_o^*,l_o^*,l_1^*$)}
%\State Initialize $N,K,M,\gbf$
\For{$n_o=1$ to $N-1$}
\For{$l_o=\lfloor \frac{KM}{N}\rfloor+1$ to $\min \{K, \lceil\frac{KM}{n_o}\rceil-1\}$}
        \State Compute $\{\abf_{n}\}$ by \eqref{equ:Twogroups2i_1} and \eqref{equ:Twogroups2i_2}.
        \State Compute $\bar{R}(n_o,l_o,l_o)$ by \eqref{equ:AverageRate}.
\EndFor
        \For{$l_o=\lfloor \frac{KM}{N}\rfloor+1$ to $K$}
        \For{$l_1=1$ to $\min\{K,\lceil \frac{KM}{n_o}\rceil -1\}$}
            \State Compute $\{\abf_{n}\}$ by \eqref{equ:Twogroup2ii_1} and \eqref{equ:Twogroup2ii_2}.
        \State Compute $\bar{R}(n_o,l_o,l_1)$ by \eqref{equ:AverageRate}.
        \EndFor
    \EndFor
    \For{$l_o=1$ to $\lfloor \frac{KM}{N}\rfloor$}
        \For{$l_1=\lceil \frac{KM}{n_o}\rceil$ to $K$}
            \State Compute $\{\abf_{n}\}$ by \eqref{equ:Twogroup2ii_1} and \eqref{equ:Twogroup2ii_2}.
        \State Compute $\bar{R}(n_o,l_o,l_1)$ by \eqref{equ:AverageRate}.
        \EndFor
    \EndFor
\EndFor
\State Compute \small $(n_o^{*},l_o^*,l_1^*)= \!\text{argmin}_{(n_o,l_o,l_1)}\bar R(n_o,l_o,l_1)$.
\State Set {\small $\bar R_{\min}=\bar R(n_o^*,l_o^*,l_1^*)$.}
\end{algorithmic}
\end{algorithm}

\begin{Remark}
In the case of two file groups,   the first possible structure of the optimal placement $\{\abf_n\}$ is described in Section \ref{subsec:ZeroFirst}: All the cache  is allocated to the first group, and the cache placement  for  files in this group is identical, \ie symmetric placement, regardless of having different file popularities among them. As mentioned earlier, this file grouping case has been considered in  \cite{Ji&Order:TIT17} and \cite{Zhang&Coded:TIT18} for a decentralized cache placement, with different methods proposed to determine the location of $n_o$.  In \cite{Ji&Order:TIT17},  for files with Zipf distribution, the selection of $n_o$ results in the performance   being a constant away from that of the optimal placement.
In \cite{Zhang&Coded:TIT18}, for an arbitrary file popularity distribution, the choice of $n_o$ results in a suboptimal caching strategy.  In contrast, we provide the optimal cache placement $\{\abf_n\}$ in Algorithm~\ref{alg:twogroup1}. The second possible structure of $\{\abf_n\}$ is shown in Section \ref{subsec:oneZeroSecond} in two possible cases, where each file in the second file group is partly cached and partly remains at the server. Different from the first structure, in this case,  coding opportunity between the two file groups is explored to minimize the average rate. We provide   Algorithm~\ref{alg:twogroup2} to determine the optimal cache placement $\{\abf_n\}$. This  placement structure has never been considered in the  literature. Depending on $(N,\pbf,M,K)$, this placement structure may lead to a higher caching gain and lower rate than the first one, as we will show in the simulation. \end{Remark}

%%%%%%%%%%%%%%%%%%%%%%%%%%%%%%%%%
\subsection{Three File Groups}\label{sec:threeGroups}

Similar to the case of two file groups, when there are three file groups under the optimal cache placement   $\{\abf_n\}$, we have three unique values among $\abf_n$'s as $\abf_{1}=\ldots=\abf_{n_o}\neq\abf_{n_o+1}=\ldots=\abf_{n_1}\neq\abf_{n_1+1}=\ldots=\abf_{N}$, for $1\le n_o<n_1\le N-1$. We use $\abf_{n_o}$, $\abf_{n_1}$ and $\abf_{n_1+1}$ to represent the three unique  placement vectors for the first, second, and third file group, respectively.
%Note that if no cache is allocated to the 3rd file group (\ie$a_{n_1+1,0}=1$), %the cache placement problem is essentially reduced to the previous two-file-group %case. Thus,
We first determine the cache placement $\abf_{n_1+1}$ in the 3rd file group  below. %The proof uses the similar technique as outlined in Proposition~\ref{Pro:TwoGroupOneDiffEle1} and is omitted.

\begin{proposition} \label{Pro:ThirdZero}
  If there are three file groups under the optimal cache placement $\{\abf_n\}$, the optimal placement vector $\abf_{n_1+1}$ for the third file group is given by $\bar\abf_{n_1+1}=\mathbf{0}$, and  $a_{n_1+1,0}=1$.
\end{proposition}
\IEEEproof
See Appendix \ref{ProofPro:ThirdZero}.
\endIEEEproof

Proposition \ref{Pro:ThirdZero} indicates that when there are three file groups under the optimal placement, all the cache will be allocated to the first two file groups; the files in the 3rd file group solely remain in the server and are not cached to any user. Following this, we only need to obtain the two unique cache placement vectors $\abf_{n_o}$ and $\abf_{n_1}$ in  the first two groups, respectively.

Note that since $\abf_{n_1}\neq\abf_{n_1+1}$, similar to \eqref{2group_a}, we have $\bar{\abf}_{n_1}\succcurlyeq_1\bar{\abf}_{n_1+1}=\mathbf{0}$ and $a_{n_1,0}<a_{n_1+1,0}=1$. As a result, the cache placement  $(\abf_{n_o}, \abf_{n_1})$ is the same as that of the two-file-group case  with $\bar{\abf}_{n_1}\succcurlyeq_1 \mathbf{0}$ for the second file group in Section~\ref{subsec:oneZeroSecond}, where $N$ is replaced by $n_1$.
Specifically, for $\bar\abf_{n_1}\succcurlyeq_1  0$,  by Propositions~\ref{Pro:TwoGroupOneDiffEle1} and  \ref{Pro:TwoGroupsTwoUnequals}, we conclude that   $\bar\abf_{n_1}$ has one nonzero element, and  $\bar\abf_{n_o}$ and $\bar\abf_{n_1}$ are different by one element. Assume $a_{n_{1},l_o}>0$, for some $l_o\in \Kc$. The  different element in $\bar\abf_{n_o}$ and $\bar\abf_{n_1}$   can be either at $l_o$ with $a_{n_o,l_o}>a_{n_1,l_o}$ (as shown in  Fig.~\ref{fig:ThreeGroup1}), or at $l_1\neq l_o$ for $l_1\in\Kc$, with  $a_{n_o,l_1}>a_{n_1,l_1}=0$ (as shown in  Fig.~\ref{fig:ThreeGroup2}).
Detailed solution for  $(\abf_{n_o}, \abf_{n_1})$ in each case can be obtained from Section \ref{subsec:oneZeroSecond}, summarized as follows:

%%%%%%%%%%%%%%%%%%%%%%%%%%
\subsubsection{When $a_{n_o,l_o}>a_{n_1,l_o}>0$} \label{case3i}
Following \eqref{equ:Twogroups2i_1} and \eqref{equ:Twogroups2i_2}, we have
\begin{align}
&\hspace*{-.5em}a_{n_o,l_o}=\frac{1}{{K \choose l_o}}, \quad a_{n_o,l}=0, \ \forall \ l\neq l_o \label{equ:Threegroups1_1}\\
&\hspace*{-1.5em}\begin{cases}\displaystyle a_{n_1,0}=\frac{1-\frac{KM}{l_on_1}}{1-\frac{n_o}{n_1}}, \
a_{n_1,l_o}\!= \frac{1}{{K \choose l_o}}\!\!\left(\!\frac{\frac{KM}{l_on_1}-\frac{n_o}{n_1}}{1-\frac{n_o}{n_1}}\!\!\right)
   \\
a_{n_1,l}= 0,  \ \forall \ l\neq 0 \ \text{or} \ l_o
\end{cases} \hspace*{-2em} \label{equ:Threegroups1_2}
\end{align}
where
$\left\lfloor\frac{KM}{n_1}\right\rfloor+1 \le  l_o \le \min \left\{K, \left\lceil\frac{KM}{n_o}\right\rceil-1\right\}$ for this case to be valid.
Note that the condition for $l_o$ can be satisfied only if $n_1 > M$. Thus, this case is possible for the optimal placement $\{\abf_n\}$ only if $n_1>M$.

%\vspace*{0.5em}
%\noindent
\subsubsection{When  $a_{n_o,l_1}>a_{n_1,l_1}=0$} \label{case3ii}
From \eqref{equ:Twogroup2ii_1} and \eqref{equ:Twogroup2ii_2}, we have
\begin{align}
&\hspace*{-1em}\begin{cases}\displaystyle a_{n_o,l_o}\!=\frac{1}{{K \choose l_o}}\frac{\frac{KM}{l_o n_1}-\frac{n_o}{n_1}}{1-\frac{n_o}{n_1}},
\
a_{n_o,l_1}=\frac{1}{{K \choose l_1}}\frac{1-\frac{KM}{l_o n_1}}{1-\frac{l_1 n_o}{l_o n_1}}
\\
a_{n_o,l}=0, \ \forall \ l\neq l_o \ \text{or} \ l_1,
\end{cases} \!\!\!\!\!\!\label{equ:Threegroup3ii_1} \\
&\hspace*{-1em}\begin{cases}\displaystyle a_{n_1,l_o}=a_{n_o,l_o}, \ \  \quad  a_{n_1,0}= \frac{1-\frac{KM}{l_oN}}{1-\frac{l_1n_o}{l_on_1}}  \\
a_{n_1,l}= 0, \forall \ l\neq 0 \ \text{or} \ l_o  \end{cases}\label{equ:Threegroup3ii_2}
\end{align}
where $l_o$ and $l_1$ need to satisfy one of the two conditions
\begin{enumerate}
\item[C1')] $l_o > KM/n_1$ and $ l_1 < KM/n_o$, or
\item[C2')] $ l_o<KM/n_1$ and $l_1 > KM/n_o$.
\end{enumerate}

\begin{figure}[t]
  \psfrag{Subgroups of a file}[][]{\quad\quad File subgroups}
  \psfrag{1st File Group}[][][0.75]{1st File Group}
  \psfrag{2nd File Group}[][][0.75]{2nd File Group}
  \psfrag{3rd File Group}[][][0.75]{3rd File Group}

  \psfrag{File 1}[][][0.8]{File $1$}
  \psfrag{$n_o$}[][][0.8]{$n_o$}
  \psfrag{$n_o+1$}[][][0.8]{$n_o+1$}
    \psfrag{$n_1$}[][][0.8]{$n_1$}
  \psfrag{$n_1+1$}[][][0.8]{$n_1+1$}
  \psfrag{$N$}[][][0.8]{$N$}

  \psfrag{$a_{1,0}$}[][][0.8]{$a_{1,0}$}
  \psfrag{$a_{1,l_o}$}[][][0.8]{$\ \ a_{1,l_o}$}
  \psfrag{$a_{1,l_1}$}[][][0.8]{$\ \ a_{1,l_1}$}
  \psfrag{$a_{1,K}$}[][][0.8]{$\ \ \ a_{1,K}$}

  \psfrag{$a_{n_o,0}$}[][][0.8]{$a_{n_o,0}$}
  \psfrag{$a_{n_o,l_o}$}[][][0.8]{$\ a_{n_o,l_o}$}
  \psfrag{$a_{n_o,l_1}$}[][][0.8]{$\ a_{n_o,l_1}$}
  \psfrag{$a_{n_o,K}$}[][][0.8]{$\ a_{n_o,K}$}

  \psfrag{$a_{n_o+1,0}$}[][][0.68]{$a_{n_o+1,0}$}
  \psfrag{$a_{n_o+1,l_o}$}[][][0.68]{$\ a_{n_o+1,l_o}$}
  \psfrag{$a_{n_o+1,l_1}$}[][][0.68]{$\ a_{n_o+1,l_1}$}
  \psfrag{$a_{n_o+1,K}$}[][][0.68]{$\ a_{n_o+1,K}$}

  \psfrag{$a_{n_1,0}$}[][][0.8]{$a_{n_1,0}$}
  \psfrag{$a_{n_1,l_o}$}[][][0.8]{$\ a_{n_1,l_o}$}
  \psfrag{$a_{n_1,l_1}$}[][][0.8]{$a_{n_1,l_1}$}
  \psfrag{$a_{n_1,K}$}[][][0.8]{$a_{n_1,K}$}

  \psfrag{$a_{n_1+1,0}$}[][][0.68]{$a_{n_1+1,0}$}
  \psfrag{$a_{n_1+1,l_o}$}[][][0.68]{$\ a_{n_1+1,l_o}$}
  \psfrag{$a_{n_1+1,l_1}$}[][][0.68]{$\ a_{n_1+1,l_1}$}
  \psfrag{$a_{n_1+1,K}$}[][][0.68]{$\ a_{n_1+1,K}$}

  \psfrag{$a_{N,0}$}[][][0.8]{$a_{N,0}$}
  \psfrag{$a_{N,l_o}$}[][][0.8]{$a_{N,l_o}$}
  \psfrag{$a_{N,l_1}$}[][][0.8]{$a_{N,l_1}$}
  \psfrag{$a_{N,K}$}[][][0.8]{$a_{N,K}$}
  \centering
  \includegraphics[scale=0.38]{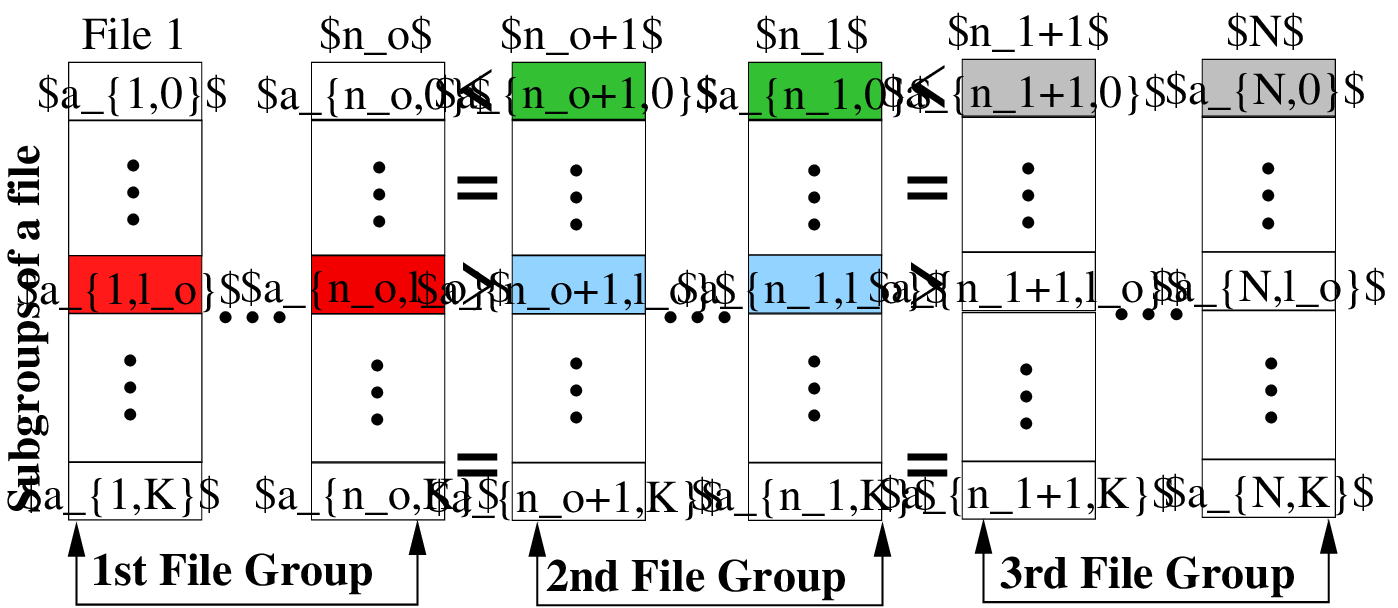}
  \renewcommand{\figurename}{Fig.}
  \caption{An example of the optimal cache placement $\{\abf_n\}$ in the case of three file groups. No cache is allocated to the 3rd file group: $a_{n_1+1,0}=1$. For $\abf_{n_o},\abf_{n_o+1}$ in the first and second groups: $1>a_{n_o+1,0}>a_{n_o,0}=0$; $a_{n_o,l_o}>a_{n_o+1,l_o}>0$,  $l_o\in\Kc$; $a_{n_o,l}=a_{n_o+1,l}=0$, $\forall l\in \Kc, l\neq l_o$.}
  \label{fig:ThreeGroup1}
\end{figure}

\begin{figure}[t]
  \psfrag{Subgroups of a file}[][][0.95]{File subgroups}
  \psfrag{1st File Group}[][][0.75]{1st File Group}
  \psfrag{2nd File Group}[][][0.75]{2nd File Group}
  \psfrag{3rd File Group}[][][0.75]{3rd File Group}

  \psfrag{File 1}[][][0.8]{File $1$}
  \psfrag{$n_o$}[][][0.8]{$n_o$}
  \psfrag{$n_o+1$}[][][0.8]{$n_o+1$}
    \psfrag{$n_1$}[][][0.8]{$n_1$}
  \psfrag{$n_1+1$}[][][0.8]{$n_1+1$}
  \psfrag{$N$}[][][0.8]{$N$}

  \psfrag{$a_{1,0}$}[][][0.8]{$a_{1,0}$}
  \psfrag{$a_{1,l_o}$}[][][0.8]{$\ \ a_{1,l_o}$}
  \psfrag{$a_{1,l_1}$}[][][0.8]{$\ \ a_{1,l_1}$}
  \psfrag{$a_{1,K}$}[][][0.8]{$\ \ \ a_{1,K}$}

  \psfrag{$a_{n_o,0}$}[][][0.8]{$a_{n_o,0}$}
  \psfrag{$a_{n_o,l_o}$}[][][0.8]{$\ a_{n_o,l_o}$}
  \psfrag{$a_{n_o,l_1}$}[][][0.8]{$\ a_{n_o,l_1}$}
  \psfrag{$a_{n_o,K}$}[][][0.8]{$\ a_{n_o,K}$}

  \psfrag{$a_{n_o+1,0}$}[][][0.68]{$a_{n_o+1,0}$}
  \psfrag{$a_{n_o+1,l_o}$}[][][0.68]{$\ a_{n_o+1,l_o}$}
  \psfrag{$a_{n_o+1,l_1}$}[][][0.68]{$\ a_{n_o+1,l_1}$}
  \psfrag{$a_{n_o+1,K}$}[][][0.68]{$\ a_{n_o+1,K}$}

  \psfrag{$a_{n_1,0}$}[][][0.8]{$a_{n_1,0}$}
  \psfrag{$a_{n_1,l_o}$}[][][0.8]{$\ a_{n_1,l_o}$}
  \psfrag{$a_{n_1,l_1}$}[][][0.8]{$a_{n_1,l_1}$}
  \psfrag{$a_{n_1,K}$}[][][0.8]{$a_{n_1,K}$}

  \psfrag{$a_{n_1+1,0}$}[][][0.68]{$a_{n_1+1,0}$}
  \psfrag{$a_{n_1+1,l_o}$}[][][0.68]{$\ a_{n_1+1,l_o}$}
  \psfrag{$a_{n_1+1,l_1}$}[][][0.68]{$\ a_{n_1+1,l_1}$}
  \psfrag{$a_{n_1+1,K}$}[][][0.68]{$\ a_{n_1+1,K}$}

  \psfrag{$a_{N,0}$}[][][0.8]{$a_{N,0}$}
  \psfrag{$a_{N,l_o}$}[][][0.8]{$a_{N,l_o}$}
  \psfrag{$a_{N,l_1}$}[][][0.8]{$a_{N,l_1}$}
  \psfrag{$a_{N,K}$}[][][0.8]{$a_{N,K}$}  \centering
  \includegraphics[scale=0.35]{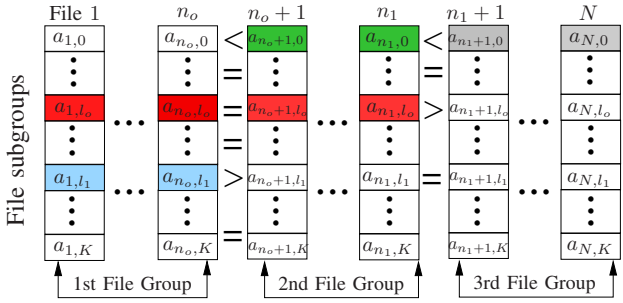}
  \renewcommand{\figurename}{Fig.}
  \caption{An example of the optimal cache placement $\{\abf_n\}$ in the case of three file groups. No cache is allocated to the 3rd file group: $a_{n_1+1,0}=1$. For $\abf_{n_o},\abf_{n_o+1}$ in the first and second groups: 1) $a_{n_o,l_1}>a_{n_o+1, l_1}=0$; 2) $a_{n_o,l_o}=a_{n_o+1,l_o}>0$; 3) $a_{n_o,l}=a_{n_o+1, l}=0$, $\forall l\in \Kc$,~$l\neq l_o,l_1$.}
  \label{fig:ThreeGroup2}
\end{figure}
Since $l_o,l_1\in\Kc$, to further analyze the above two conditions for $l_o$ and $l_1$, we note that
\begin{itemize}
\item If $n_o<n_1\le M$: neither C1') nor C2') can be satisfied;
\item If $n_o\le M < n_1$: only C1') can be satisfied;
\item If $M<n_o < n_1$: both C1') and C2') are possible.
\end{itemize}
As a result,  Case 2) is only possible for the optimal placement $\{\abf_n\}$  if $n_1>M$.

The structure of $\{\abf_n\}$ in Cases 1) and 2) are illustrated in Figs.~\ref{fig:ThreeGroup1} and ~\ref{fig:ThreeGroup2}, respectively, where the colored blocks indicate the nonzero elements in $\abf_n$.

\begin{Remark}\label{remark3}
From  Cases 1) and 2) above, we conclude that if the optimal placement results in three file groups, we must have $n_1>M$. This result is consistent with our intuition: By Proposition~\ref{Pro:ThirdZero}, all the cache is allocated to the first two file groups. To maximally use the cache, the files to be cached (in the first two groups) must be no less than $M$ files.
\end{Remark}

Based on the above discussion, for the case of three file groups, given ($n_o, n_1, l_o,l_1$), the solution  $\{\abf_n\}$ is obtained in closed-form, and so the average rate $\bar{R}$ in {\bf P2} can be computed by \eqref{equ:AverageRate} as a function of $(n_o,n_{1,}l_o,l_1)$. Again, we can  search over all possible values of $n_1\in\{M+1,N-1\}$, $n_o\in\{1,\ldots,n_1-1\}$,  and $l_o,l_1\in \Kc$ within the range specified in Cases 1) and 2), to obtain the best tuple $(n_o,n_{1,}l_o,l_1)$  that gives minimum  $\bar{R}$. Algorithm~\ref{alg:threeGroups} summarizes the steps to obtain the best placement solution  $\{\abf_n\}$ for three file groups. It uses Algorithm \ref{alg:twogroup2} to obtain the best tuple $(n_o,l_o,l_1)$ in the two-file-group subproblem, for each $n_1\in\{M+1,\ldots,N-1\}$.
The algorithm simply computes $\bar{R}$ for different $(n_o,n_1,l_o,l_1)$ using the closed-form expression in  \eqref{equ:AverageRate}  for  at most $(N-1)(N-M-1)K^2/2$ times in the worst case (depending on the values of $(N,M,K)$) . They can be computed efficiently in parallel.

\begin{algorithm}[t]
\caption{The Cache Placement for Three File Groups}\label{alg:threeGroups}
\begin{algorithmic}[1]
\Require
    {$K$,  $M$,  $N$, and  $\pbf$}
\Ensure
    {($\bar R_{\min}$, $n_o^*$, $n_1^*$, $l_o^*$, $l_1^*$)}
%\State Initialize $N,K,M,\gbf$
\For{$n_1=M$ to $N-1$}
    \State $\bar R_1(n_o, n_1, l_o, l_1) =$\\ \hspace{5em} $\text{Algorithm}~\ref{alg:twogroup2}(K,M,n_1,[p_1,\ldots,p_{n_1}]^T)$;
    \State $\bar R_2(n_1) =  \sum_{n=n_1+1}^{N}g_{n,0}$;
    \State Compute $\bar R(n_o, n_1, l_o, l_1) = \bar R_1 + \bar R_2$
\EndFor
\State Compute {\small $(n_o^{*},n_1^*,l_o^*,l_1^*)= \textrm{argmin}_{ n_o,n_1,l_o,l_1}\bar R(n_o,n_1,l_o,l_1)$};
\State Set $\bar R_{\min}=\bar R(n_o^*,n_1^*,l_o^*,l_1^*)$.
\end{algorithmic}
\end{algorithm}
\begin{algorithm}[t]
\caption{The Optimal Cache Placement Solution for {\bf P1}}\label{alg:placement}
\begin{algorithmic}[1]
\Require
    {$K$,  $M$,  $N$, and  $\pbf$}
\Ensure
    {$\bar{R}_{\min}$, $\{\abf_1,\ldots,\abf_N\}$}
%\State Initialize $N,K,M,\gbf$
\State Run Algorithms\ \ref{alg:twogroup1}, \ref{alg:twogroup2} and \ref{alg:threeGroups}.
\State Find the minimum output $\bar{R}_{\min}$ among the outputs of Algorithms \ref{alg:twogroup1}--\ref{alg:threeGroups}.
\State Set the corresponding placement $\{\abf_1,\ldots,\abf_N\}$ for $\bar{R}_{\min}$ as the optimal $\{\abf_1,\ldots,\abf_N\}$.
%\Return $\bar R_{\min}$ and $\{\abf_1,\ldots,\abf_N\}$.
%\Return
\end{algorithmic}
\end{algorithm}

\begin{Remark}
We point out that there is no  three-file-group caching scheme proposed for the CCS in the literature. Only  \cite{Zhang&Coded:TIT18} has considered  adding   a specific three-file-group case heuristically as part of a mixed caching scheme, where  the second file group contains only one file.  However, uncoded caching is used for the case of three file groups, \ie the content delivery is  uncoded, and the case is  used for very rare occasions.  In the simulation, we will show that the three-file-group cache placement for coded caching is optimal and outperforms the two-group strategy even for files with Zipf distribution.
\end{Remark}

%%%%%%%%%%%%%%%%%%%%%%%%%%%%%%%%%
\subsection{The Optimal Cache Placement Solution}\label{subsubsec:opt_placement}
 By Theorem~\ref{The3Groups},
 the optimal cache placement problem {\bf P1} (or {\bf P2}) is reduced to three subproblems, \ie one, two, or three file groups, respectively.  The possible structure of the optimal cache placement in each subproblem  is given in Sections~\ref{sec:oneGroup} to \ref{sec:threeGroups}. These results lead to a simple algorithm to obtain the optimal  placement solution $\{\abf_n\}$ for {\bf P1}: Each file-group case returns the candidate  optimal solution $\{\abf_n\}$ with the minimum  $\bar{R}$ for this subproblem. The optimal $\{\abf_n\}$ can then be obtained by taking the one that gives the minimum $\bar{R}$ among the three subproblems. The details are summarized in Algorithm \ref{alg:placement}. It uses Algorithms \ref{alg:twogroup1}--\ref{alg:threeGroups} and selects $\{\abf_n\}$ that returns the minimum $\bar R$ as the optimal solution.
Again, we point out obtaining the optimal $\{\abf_n\}$  in Algorithm \ref{alg:placement} requires minimum complexity.  Algorithms  \ref{alg:twogroup1}--\ref{alg:threeGroups} each involves computing a closed-form expression of $\bar{R}$  multiple times, and all can be done in parallel. In total, $\bar{R}$ is computed for at most  $(N-1)(N-M+1)K^2/2+N$ times in the worst case.\footnote{
Under the optimal placement, files with the same popularity have identical placement, \ie $a_{n,l}=a_{n',l}$, $\forall l$, if $p_n=p_n'$. This means that the files with the same popularity are  in the same file group (\eg a single file group for files with uniform popularity). This may further reduce the set of candidate solutions in Algorithms  \ref{alg:twogroup1}--\ref{alg:threeGroups} by only considering possible values of $n_o$ (and $n_1$) only for  $p_{n_o}>p_{n_o+1}$ (and $p_{n_1}>p_{n_1+1}$).}

How to determine the file groups depends on $(\pbf, N, K, M)$. Although Sections~\ref{sec:oneGroup} to \ref{sec:threeGroups} provide the possible structure of the optimal cache placement in three file grouping cases, analytically determining the final optimal file grouping, \ie the number of file groups and the group partition ($n_o$ for two groups, and $(n_o,n_1)$ for three groups), is still challenging. The same for the location of nonzero element(s) $l_o$ (and $l_1$) in $\abf_n$, \ie the choice of cache subgroup(s) for subfiles. They depend on the file popularity distribution $\pbf$, {the number of users $K$} and the relative cache size to the database size ($M$ vs. $N$). Our proposed Algorithm~\ref{alg:placement} that combines Algorithms~\ref{alg:twogroup1}--\ref{alg:threeGroups} provides a simple and efficient method to obtain the optimal file grouping.  Using the obtained file group structures, Algorithms~\ref{alg:twogroup1}--\ref{alg:threeGroups} significantly simplify the solving of {\bf P1}, by providing a set of candidate solutions in closed-form in each case.

%\begin{Remark}\label{remark1}
\subsection{{Discussion on the Optimal File Group Structure}}\label{sec:discussion}
{The result in Theorem 1 of having at most three file groups in the optimal cache placement for the CCS}, regardless of file distribution $\pbf$, is somewhat surprising. Based on the results obtained in Sections~\ref{sec:oneGroup} to \ref{sec:threeGroups}, we provide some insights into  the optimal file group structure. We can recognize the three file groups as three categories of ``most popular," ``moderately popular," and ``non-popular" files. Regardless of file popularity distribution $\pbf$, the caching method only distinguishes files by one of these three categories. The three categories reflect the caching strategies:
From the structure of optimal $\{\abf_n\}$  obtained in Sections~\ref{sec:oneGroup} to \ref{sec:threeGroups}, the optimal caching strategy is to 1) cache all subfiles of the ``most popular" files (among $K$ users); 2) for the deemed ``moderately popular" files,  cache only a portion of each file, and leave the rest  solely at the server; 3) if there are ``non-popular" files, they are not cached but only stored in the server.

%For the nontrivial case of $M>0$, the category of ``most popular" files always exists. Whether files are deemed as the other two categories depend on $\pbf$, $KM/N$. There may be one or two file groups under the optimal cache placement. In the case of one group, it  means that all files are deemed ``most popular;" and in the case of two groups, it means there are  files deemed in either ``moderately popular" or  ``non-popular" file groups. 

{Note that a file belongs to which category is a relative notion:   the mapping of files into these three categories, \ie file grouping,  depends on the file popularity distribution $\pbf$ and  the  ratio of  global cache size to the database size $KM/N$. 
To further understand the file  grouping phenomenon and the case of three file groups, we provide numerical examples  in Section \ref{sec:sim:placement}  through Tables~\ref{solutionM1}--\ref{solutionM7} to show how the file group structure changes   (see Section \ref{sec:sim:placement} for the detailed discussion).  As  $M/N$ increases (\eg from $10\%$ to $80\%$),  we observe that the optimal number of  file groups changes as follows: $2\to 3\to 2 \to 3 \to 1$. Intuitively, increasing the cache memory allows more files to be cached. As a result, a  file deemed ``non-popular" for small cache size may be deemed ``most popular" for large cache size. Thus, when $M/N$ increases, more files are shifted from the ``non-popular" group  (only stored in the server) to the ``most popular" group  (all cached), with fewer files in the ``non-popular" group. During this transition, the ``moderately popular" file group (partly cached) appears, as the cache size is large enough to partly store some file but not all of it (among users). This explains why and when three file groups become  optimal for the cache placement.}

{The existing two-file-group schemes proposed in \cite{Ji&Order:TIT17,Zhang&Coded:TIT18} have ``non-popular" and ``most popular" groups and use a suboptimal strategy to decide the file groups. They can be viewed as reflections of the two-file-group scenario. However, these two-file-group schemes cannot capture the ``moderately popular" group during the  transition stage mentioned above. In contrast, the optimal  solution we obtain captures all possible file groups, which provides the highest resolution in determining the optimal cache placement, leading to  the minimum rate.
}

%%%%%%%%%%%%%%%%%%%%%%%%%%%%%%%%%%%%%%%%%%%%%%%%
\section{ Converse Bound}\label{sec:bound}
In this section, we show that  the structure of the optimal cache placement solution for $\textrm{\bf P1}$ obtained earlier can be used to obtain a  tighter information-theoretic lower bound on the  average rate $\bar{R}$  for any coded caching scheme (with uncoded or coded cache placement), under arbitrary file popularity. This converse bound is obtained using a genie-based method. Some existing works~\cite{Niesen&Maddah-Ali:TIT2017,Ji&Order:TIT17,Zhang&Coded:TIT18} have used this genie-based method to derive the lower bounds on the  average rate with different tightness. This genie-based method constructs a virtual system, where only a group of popular files are delivered to the users   via the shared link, and the rest (unpopular) files are delivered by a genie instead of using the shared link.
Furthermore, the virtual system treats this group of popular files as if they have  uniform popularity, leading to the symmetric cache placement strategy with the same placement for all these files.
The average rate of the original system under any coded caching scheme is shown to be lower bounded by that of this virtual system~\cite{Niesen&Maddah-Ali:TIT2015}.

In deriving a lower bound using the genie-based method, the determination of the group of popular files plays an important role in the tightness of the  bound. Let $p^{\prime}$ be the probability threshold  to decide the group of popular files, where file $W_n$ belongs to this group if $p_n\geq p^{\prime}$.
Let $N_{p'}$ denote the number of popular files in the group. The general result for the lower bound shows that, for $K$ users requesting files independently, the average rate is lower bounded by \cite{Zhang&Coded:TIT18}
\begin{align}\label{R_lb}
\bar{R} \ge \bar R^{\text{lb}}=\frac{1}{11}Kp^{\prime}(N_{p'}-M).
\end{align}

%By proposing different suboptimal cache placement structures,
  Heuristical methods are used to decide  the group of popular files to derive the converse bounds. In \cite{Ji&Order:TIT17}, specific for the Zipf distribution, the choice of  $N_{p'}$ is proposed for different Zipf parameter values, file sizes, and cache sizes.
In~\cite{Zhang&Coded:TIT18}, the value of $p'$ is proposed for an arbitrary file popularity distribution.
To tighten the bound further, a  file merging approach is proposed in~\cite{Zhang&Coded:TIT18}: Those files not belonging to the group of popular files,
 but deemed moderately popular, are merged into new virtual files to be included in the group of popular files.
Specifically, by the definition of $N_{p'}$, we have $p_{N_{p'}+1}<p^{\prime}$.
 From file $W_{N_{p'}+1}$ and afterwards, subsequent files are merged into a new virtual file until the accumulated popularity of these merged files exceeds $p^{\prime}$.  The procedure repeats until all the rest files are considered. Let
 $N_{p'}^\text{m}$ denote the number of virtual files generated by the merging procedure.
 With these additional  virtual files, there are
$N_{p'}+N_{p'}^\text{m}$ popular files. Using \eqref{R_lb}, ~\cite{Zhang&Coded:TIT18} shows a tighter lower bound given by
\begin{align}\label{equ_lowerbound_merge}
 \bar R^{\text{lb}}=\frac{1}{11}Kp^{\prime}(N_{p'}+N_{p'}^\text{m}-M),
\end{align}
and the  number of virtual files is
$N_{p'}^\text{m}=\lfloor\frac{\sum_{n>N_{p'}}p_n}{2/p^{\prime}}+\frac{1}{2}\rfloor$.
 The file merging approach allows some moderately popular files to be considered in deriving the converse bound.
As a result, the  bound in \eqref{equ_lowerbound_merge} is by far the tightest converse bound.

The value of $p'$ for the converse bound in \eqref{equ_lowerbound_merge} obtained in ~\cite{Zhang&Coded:TIT18} is determined by combining a heuristic method and the exhaustive search.
 The method  sets $p'=p_1\triangleq\frac{1}{K\max\{3,M\}}$,
which results in  $N_{p_1}$ popular files and $N_{p_1}^\text{m}$ virtual files. To avoid trivial negative lower bound in \eqref{equ_lowerbound_merge}, when $N_{p_1}+N_{p_1}^\text{m}<M$, the exhaustive search of  $p'$ ($N_{p'}$) is used  by searching over the rest of files with popularity less than $p_1$, \ie $\{W_n: N_{p_1}+1\le n\le N\}$. As a result, the converse bound is given by \cite{Zhang&Coded:TIT18}
\begin{align}\label{equ_lowerbound_zhang}
  \bar R^{\text{lb}}= &\frac{1}{11}\max\left\{\frac{1}{\max\{3,M\}}(N_{p_1}+N_{p_1}^\text{m}-M), \right. \nn\\
  &\left. \max_{ N_{p_1}+1 \le  n \le N}K p_n(N_{p_n}+N_{p_n}^\text{m}-M) \right\}    %(N_1+N_2-M).
\end{align}
where the second term provides a possible improved converse bound through the exhaustive search for $N_p' \in\{N_{p_1}+1,\ldots,N\}$.

 Interestingly, the use of popular files to derive the lower bound echoes the structure of the optimal cache placement solution for \textbf{P1}. As discussed in Section~\ref{subsubsec:opt_placement},  the  $n_o$ files in the first file group are the most popular files for caching.
Based on this observation,  we determine  $N_{p'}$  by the optimal cache placement for the CCS. The group of the most popular files  is obtained from Algorithm~\ref{alg:placement} with size $n_o$, with the corresponding file popularity threshold  set as $p^{\prime}=p_{n_o}$.
 The number of virtual files is $N_{p_{n_o}}^\text{m}$ accordingly. Then, we obtain the lower bound  $\bar{R}^\text{lb}$ as follows.
\begin{proposition}
  Let $n_o$ be the number of files in the first file group by the optimal cache placement solution for $\textrm{\bf P1}$. The  average delivery rate is lower bounded by
  \begin{align}\label{equ_lb_opt}
    \bar{R}\ge \bar R^{\text{lb}}= \frac{1}{11}Kp_{n_o}(n_o+N_{p_{n_o}}^\text{m}-M).
  \end{align}
\end{proposition}

We point out that the difference of $\bar{R}^{\text{lb}}$ in \eqref{equ_lb_opt} from the existing methods  \cite{Ji&Order:TIT17,Zhang&Coded:TIT18} is that, instead of determining the popular files heuristically or through an exhaustive search, we obtain  the number of most popular files  $n_o$ from the optimal file group structure in the cache placement optimization.  In the simulation, we show that the  lower bound in \eqref{equ_lb_opt} is tighter than the existing ones, especially for a smaller cache size when  the average rate is more sensitive to cache placement. This shows that using the file groups given by the optimal cache placement  for the CCS provides a more accurate method in determining the popular files than existing methods.

%%%%%%%%%%%%%%%%%%%%%%%%%%%%%%%%%%%%%%%%%%%%%%%%
\section{Subpacketization Upper Bound}\label{sec:subpacket}
The subpacketization level, \ie the number of  subfiles in each file required for caching, is an important issue for the practical implementation of coded caching. Since the optimal cache placement has not been characterized before, there is no clear quantification of the number of subfiles generated by  the CCS. In this section,
we explore the properties in the optimal cache placement
solution for \textrm{\bf P1} to characterize the subpacketization structure and derive an upper bound on the subpacketization level  under the optimal cache placement, for any  file popularity distribution $\pbf$ and memory size $M$.

Recall from Section~\ref{III.A} that each file can be partitioned into  $2^K$ subfiles, which are divided into $K+1$ file subgroups $\Wc_n^l$, $l\in \Kc\cup\{0\}$. There are   ${K \choose l}$ subfiles  in $\Wc_n^l$,  each with size   $a_{n,l}$. They will be stored in corresponding user subsets with size $l$, provided that $a_{n,l}>0$.
The subpacketization level  $L_n$ of file  $n$ is directly related to its placement vector $\abf_{n}$ as
$L_n=\sum_{l\in\Kc\cup\{0\}:a_{n,l}>0}\binom{K}{l}$.
Based on the  structure of the optimal cache placement  $\{\abf_n\}$ presented in Section~\ref{sec:Placement},
it is straightforward to conclude the following property of $\abf_n$.
\begin{corollary}\label{cor:The2subfiles}
   For $N$ files with any  file popularity distribution $\pbf$, the optimal cache placement  $\abf_{n}$  of any file $n$ for \textrm{\bf P1} has at most two nonzero elements.
 \end{corollary}

Following this property,  we bound the worst-case maximum subpacketization level, defined by $L^{\text{max}}=\max_n L_n$, for the CCS.
\begin{proposition} \label{prop:subpacket}
  For given $(N,\pbf,M,K)$, the maximum subpacketization level  $L^\text{max}$  under the optimal cache placement for the CCS is bounded by
\begin{align}\label{L_bd}
L^{\max}\le \binom{K}{\lfloor K/2 \rfloor}+\binom{K}{\lfloor K/2 \rfloor+1}\leq \sqrt{\frac{8}{\pi}}e^{\frac{1}{12K}}\frac{2^K}{\sqrt{K}}.
\end{align}
\end{proposition}
\IEEEproof
From Corollary \ref{cor:The2subfiles}, by the optimal cache placement solution, the subfiles of any file belong to at most two file subgroups of different sizes.
There are $\binom{K}{l}$ subfiles need to be cached into the user subsets with size $l\in\Kc\cup\{0\}$.
Then, for $l=\lfloor K/2 \rfloor$ and $\lfloor K/2 \rfloor+1$,  the number of subfiles is the highest. Consequently, we have $L^{\max}\le \binom{K}{\lfloor K/2 \rfloor}+\binom{K}{\lfloor K/2 \rfloor+1}$.
Based on the Stirling's approximation \cite{Feller:Stirling68}, we have
\begin{align*}
\sqrt{2\pi K}\left(\frac{K}{e}\right)^{K} \le K!\leq \sqrt{2\pi K} \left(\frac{K}{e}\right)^{K}\!\!e^{\frac{1}{12K}},
%,\;\; \;\;\left(\frac{K}{2}\right)!\geq \sqrt{2\pi}\left(\frac{K}{2}\right)^{\frac{K}{2}+\frac{1}{2}}\!\!e^{-\frac{K}{2}}.
\end{align*}
where the bounds become tight as $K$ increases.
Assuming $K=2m$, $m \in \mathbb{N}^+$, we have%\end{align}
\begin{align}
\binom{K}{\frac{K}{2}}=\frac{K!}{\frac{K}{2}!\cdot\frac{K}{2}!}\leq\frac{\sqrt{2\pi} e^{\frac{1}{12K}}K^{K+\frac{1}{2}}\!e^{-K}}{{2\pi}\left(\frac{K}{2}\right)^{K+1}e^{-K}}
\leq\sqrt{\frac{2}{\pi}}e^{\frac{1}{12K}}\frac{2^K}{\sqrt{K}},\nn
\end{align}
and we have \eqref{L_bd}.
\endIEEEproof

Proposition~\ref{prop:subpacket} indicates that the maximum number of subfiles in the worst-case grows as $\Oc(2^K/\sqrt{K})$. The actual subpacketization level of a file group depends on the location of nonzero element $l_o$ ($l_1$) in $\abf_n$. Although we cannot explicitly obtain $l_o$ ($l_1$) for the optimal placement, in general, for given $K$, it is a function of  the   cache size relative to the database size $M/N$. Recall that for smaller $l$, subfiles in $\Wc_n^l$  are cached to smaller user subsets $\Sc$'s ($|\Sc|=l$), and vice versa. Intuitively, this means that the location $l_o$ ($l_1$) of the nonzero element tends to be smaller for smaller cache size and becomes larger as $M/N$ increase. This intuition is\ confirmed by experiments. In the simulation, we show that, depending on $M/N$, the actual subpacketization level of the optimal cache placement  typically can be much less than the  upper bound in \eqref{L_bd}.
%Finally, note that in Proposition~\ref{prop:subpacket}, we focus on the scaling behavior of the worst-case subpacketization level.  A tighter upper bound on $L^{\max}$ is provided in our conference paper~\cite{Deng2019Subpacketization} which has the same scaling behavior.   %\begin{Remark}

\begin{Remark}
The tradeoff between the average rate and the subpacketization level has been studied  in \cite{Jin&Cui:Arxiv2018} via a numerical search over different subpacketization levels.
We point out that the upper bound  in Proposition \ref{prop:subpacket} provides the exact  subpacketization level, for which   increasing  it further no longer leads to a rate reduction.
\end{Remark}

\section{Simulation Results}\label{sec:simu}
In this section, we evaluate the performance of the optimal cache placement by the proposed algorithm  and  the corresponding subpacketization level for different system setups. Further, we also evaluate the  information-theoretic lower bound based on the file grouping strategy
in the optimal cache placement solution.
%%%%%%%%%%%%%%%%%%%%%%%
\subsection{The Optimal Cache Placement}\label{sec:sim:placement}
\renewcommand{\arraystretch}{1.05}
\begin{table}[t]
\centering\caption{Cache placement matrix for $K=7$,
 $N=9$, $\theta=1.5$, $M=1$.}\label{solutionM1}\resizebox{8.75cm}{!}{
\begin{tabular}{|p{1mm}<{\centering}|p{6mm}<{\centering}|p{6mm}<{\centering}|p{6mm}<{\centering}|p{6mm}<{\centering}|p{6mm}<{\centering}|p{6mm}<{\centering}|p{6mm}<{\centering}|p{6mm}<{\centering}|p{6mm}<{\centering}|p{6mm}<{\centering}} %{|c|c|c|c|c|c|c|c|c|c|c}
\hline
\multirow{2}{*}{$l$}&\multicolumn{9}{c|}{Cache placement vector of each file}\\ \cline{2-10}
          &$\abf_1$&$\abf_2$&$\abf_3$&$\abf_4$&$\abf_5$&$\abf_6$&$\abf_7$&$\abf_8$&$\abf_9$\\
          \hline\hline
 $0$    &0     &0     &0     &1.0000&1.0000&1.0000&1.0000&1.0000&1.0000\\
 $1$    &0     &0     &0     & 0    &0     &0     &0     &0     &0     \\
 $2$    &0.0317&0.0317&0.0317&0     &0     &0     &0     &0     &0     \\
 $3$    &0.0095&0.0095&0.0095&0     &0     &0     &0     &0     &0     \\
 $4$    &0     &0     &0     &0     &0     &0     &0     &0     &0     \\
 $5$    &0     &0     &0     &0     &0     &0     &0     &0     &0     \\
 $6$    &0     &0     &0     & 0    &0     &0     &0     &0     &0     \\
 $7$    &0     &0     &0     & 0    &0     &0     &0     &0     &0     \\
\hline
\end{tabular}}
%\end{table}
\vspace{3mm}
%\renewcommand{\arraystretch}{1.05}
%\begin{table}[t]
\centering\caption{Cache placement matrix for $K=7$,
$N=9$, $\theta=1.5$, $M=2.5$.}\label{solutionM2.5}\resizebox{8.75cm}{!}{
\begin{tabular}{|p{1mm}<{\centering}|p{6mm}<{\centering}|p{6mm}<{\centering}|p{6mm}<{\centering}|p{6mm}<{\centering}|p{6mm}<{\centering}|p{6mm}<{\centering}|p{6mm}<{\centering}|p{6mm}<{\centering}|p{6mm}<{\centering}|p{6mm}<{\centering}} %{|c|c|c|c|c|c|c|c|c|c|c}
\hline
\multirow{2}{*}{$l$}&\multicolumn{9}{c|}{Cache placement vector of each file}\\ \cline{2-10}
          &$\abf_1$&$\abf_2$&$\abf_3$&$\abf_4$&$\abf_5$&$\abf_6$&$\abf_7$&$\abf_8$&$\abf_9$\\
          \hline\hline
 $0$    &0     &0     &0     &0     &0.2500&0.2500&1.0000&1.0000&1.0000\\
 $1$    &0     &0     &0     &0     &0     &0     &0     &0     &0     \\
 $2$    &0     &0     &0     &0     &0     &0     &0     &0     &0     \\
 $3$    &0.0214&0.0214&0.0214&0.0214&0.0214&0.0214&0     &0     &0     \\
 $4$    &0.0071&0.0071&0.0071&0.0071&0     &0     &0     &0     &0     \\
 $5$    &0     &0     &0     &0     &0     &0     &0     &0     &0     \\
 $6$    &0     &0     &0     & 0    &0     &0     &0     &0     &0     \\
 $7$    &0     &0     &0     & 0    &0     &0     &0     &0     &0     \\
\hline
\end{tabular}}
%\end{table}
\vspace{3mm}
\renewcommand{\arraystretch}{1.05}
%\begin{table}
\centering\caption{Cache placement matrix for $K=7$,
$N=9$, $\theta=1.5$, $M=4$.}\label{solutionM4}\resizebox{8.75cm}{!}{
\begin{tabular}{|p{1.0mm}<{\centering}|p{6mm}<{\centering}|p{6mm}<{\centering}|p{6mm}<{\centering}|p{6mm}<{\centering}|p{6mm}<{\centering}|p{6mm}<{\centering}|p{6mm}<{\centering}|p{6mm}<{\centering}|p{6mm}<{\centering}|p{6mm}<{\centering}} %{|c|c|c|c|c|c|c|c|c|c|c}
\hline
\multirow{2}{*}{$l$}&\multicolumn{9}{c|}{Cache placement vector of each file}\\ \cline{2-10}
          &$\abf_1$&$\abf_2$&$\abf_3$&$\abf_4$&$\abf_5$&$\abf_6$&$\abf_7$&$\abf_8$&$\abf_9$\\
          \hline\hline
 $0$    &0     &0     &0     &0     &0     &0     &0     &1.0000&1.0000\\
 $1$    &0     &0     &0     & 0    &0     &0     &0     &0     &0     \\
 $2$    &0     &0     &0     &0     &0     &0     &0     &0     &0     \\
 $3$    &0     &0     &0     &0     &0     &0     &0     &0     &0     \\
 $4$    &0.0286&0.0286&0.0286&0.0286&0.0286&0.0286&0.0286&0     &0     \\
 $5$    &0     &0     &0     &0     &0     &0     &0     &0     &0     \\
 $6$    &0     &0     &0     & 0    &0     &0     &0     &0     &0     \\
 $7$    &0     &0     &0     & 0    &0     &0     &0     &0     &0     \\
\hline
\end{tabular}}
%\end{table}
\vspace{3mm}
%\renewcommand{\arraystretch}{1.05}
%\begin{table}[t]
\centering\caption{Cache placement matrix for $K=7$,
$N=9$, $\theta=1.5$, $M=5.5$.}\label{solutionM5.5}\resizebox{8.75cm}{!}{
\begin{tabular}{|p{1mm}<{\centering}|p{6mm}<{\centering}|p{6mm}<{\centering}|p{6mm}<{\centering}|p{6mm}<{\centering}|p{6mm}<{\centering}|p{6mm}<{\centering}|p{6mm}<{\centering}|p{6mm}<{\centering}|p{6mm}<{\centering}|p{6mm}<{\centering}} %{|c|c|c|c|c|c|c|c|c|c|c}
\hline
\multirow{2}{*}{$l$}&\multicolumn{9}{c|}{Cache placement vector of each file}\\ \cline{2-10}
          &$\abf_1$&$\abf_2$&$\abf_3$&$\abf_4$&$\abf_5$&$\abf_6$&$\abf_7$&$\abf_8$&$\abf_9$\\
          \hline\hline
 $0$    &0     &0     &0     &0     &0     &0     &0     &0.3000&1.0000\\
 $1$    &0     &0     &0     & 0    &0     &0     &0     &0     &0     \\
 $2$    &0     &0     &0     &0     &0     &0     &0     &0     &0     \\
 $3$    &0     &0     &0     &0     &0     &0     &0     &0     &0     \\
 $4$    &0     &0     &0     &0     &0     &0     &0     &0     &0     \\
 $5$    &0.0476&0.0476&0.0476&0.0476&0.0476&0.0476&0.0476&0.0333&0     \\
 $6$    &0     &0     &0     & 0    &0     &0     &0     &0     &0     \\
 $7$    &0     &0     &0     & 0    &0     &0     &0     &0     &0     \\
\hline
\end{tabular}}
\end{table}

We first verify   the structure of the optimal cache placement solution for the  CCS obtained in Section~\ref{sec:Placement}. To do so, we obtain  the placement solution $\{\abf_n\}$ by Algorithm~\ref{alg:placement} and verify that they match the optimal $\{\abf_n\}$ obtained by numerically solving \textbf{P1}.  For example, we generate user random demands using Zipf distribution, where file $n$ is requested  with probability $p_{n}=\frac{n^{-\theta}}{\sum_{i=1}^{N}i^{-\theta}}$ , with $\theta>0$ being the Zipf parameter.
For $N=9$, $\theta=1.5$, and $K=7$, Tables~\ref{solutionM1} - \ref{solutionM7} show  the optimal $\{\abf_n\}$  for cache size $M=1, 2.5, 4, 5.5, 6, 7$, respectively.
They cover the possible cases of the optimal cache placement structure  discussed in Section~\ref{sec:Placement}. As  the cache size increases from small to large, different file groups and subfile partition strategies under the optimal placement solution can be observed.
In all these results, the cache placement vectors $\{\abf_n\}$ have at most two nonzero elements, as stated in Corollary~\ref{cor:The2subfiles}.

Tables~\ref{solutionM1}  shows the optimal cache
placement solution $\{\abf_n\}$ for $M=1$. There are two file groups under the optimal solution, as in the case discussed in Section~\ref{subsec:ZeroFirst} (Fig.~\ref{fig:TwoGroup1}).
They are deemed ``most popular" and ``non-popular" files. The placement vector of the first file group has two nonzero elements (\eg  file $W_1$ is partitioned into two subfile groups $\Wc_1^2$ and $\Wc_1^3$, containing subfiles of size $0.0317$ and $0.0095$, respectively), and  the  files in the second group are  only stored at the server.

As $M$ is increased to $2.5$, Tables~\ref{solutionM2.5} shows that the optimal placement divides files into three file groups, verifying the structure  of the optimal $\{\abf_n\}$ described in   Section~\ref{case3ii} and illustrated in Fig.~\ref{fig:ThreeGroup2}. The ``moderately popular" file group ($\{W_5,W_6\}$) is included in this case,  for which  the increased cache size allows more room to cache a portion of these files, while leaving the rest portion at the server. Between the first two groups, we observe that the sub-placement vectors $\bar{\abf}_n$'s are only different by one element.

When $M$ is further increased to $4$, we observe from Table~\ref{solutionM4} that the placement results in two file groups, similar to that for $M=1$. However, compared to $M=1$, larger cache memory allows more files to be considered in the ``most popular" file group to be cached. For these files,  $\abf_n$ has only one nonzero element, indicating they are all partitioned into subfiles of equal length.

For $M=5.5$ in Table~\ref{solutionM5.5},   the files are divided into three groups, where file $W_8$ is now considered ``moderately popular" and partly cached, instead of ``non-popular"  as in the case of $M=4$. Table~\ref{solutionM5.5} is  the case described in  Fig.~\ref{fig:ThreeGroup1} of Section~\ref{case3i}. As we keep increasing $M$, we see from Tables~\ref{solutionM6} that for $M=6$,  the result is as described in Fig.~\ref{fig:TwoGroup2}, where files are considered either ``most popular" or ``moderately popular" and are stored among users accordingly. For $M=7$, Table~\ref{solutionM7}  shows that when there is enough cache  at users, all files are considered ``most popular" with identical cache placement as discussed in Section~\ref{sec:oneGroup}. This single file group resembles  the placement under uniform file popularity.

From $M=1$ to $M=7$, we notice that the location of the nonzero element in $\abf_n$ (the value of $l_o$ and $l_1$) is increasing. This indicates that as $M$ increases,  each subfile is stored into a larger user subset. This trend confirms our intuition that the optimal $l_o$ ($l_1$) increases as more cache memory is added.

Note that the optimal placement solutions in Tables~\ref{solutionM2.5},  \ref{solutionM5.5}, and \ref{solutionM6} show three or two file groups that have not been considered in the existing suboptimal   schemes. For example, in  \cite{Ji&Order:TIT17},
only two file groups are considered, with the second group  of files  kept at the server. As a result, these existing
schemes cannot always guarantee the minimum rate.

%$\vspace{3mm}
\renewcommand{\arraystretch}{1.05}
\begin{table}[t]
\centering\caption{Cache placement matrix for $K=7$,
$N=9$, $\theta=1.5$, $M=6$.}\label{solutionM6}\resizebox{8.75cm}{!}{
\begin{tabular}{|p{1mm}<{\centering}|p{6mm}<{\centering}|p{6mm}<{\centering}|p{6mm}<{\centering}|p{6mm}<{\centering}|p{6mm}<{\centering}|p{6mm}<{\centering}|p{6mm}<{\centering}|p{6mm}<{\centering}|p{6mm}<{\centering}|p{6mm}<{\centering}} %{|c|c|c|c|c|c|c|c|c|c|c}
\hline
\multirow{2}{*}{$l$}&\multicolumn{9}{c|}{Cache placement vector of each file}\\ \cline{2-10}
          &$\abf_1$&$\abf_2$&$\abf_3$&$\abf_4$&$\abf_5$&$\abf_6$&$\abf_7$&$\abf_8$&$\abf_9$\\
          \hline\hline
 $0$    &0     &0     &0     &0     &0     &0     &0     &0     &0.6000\\
 $1$    &0     &0     &0     & 0    &0     &0     &0     &0     &0     \\
 $2$    &0     &0     &0     &0     &0     &0     &0     &0     &0     \\
 $3$    &0     &0     &0     &0     &0     &0     &0     &0     &0     \\
 $4$    &0     &0     &0     &0     &0     &0     &0     &0     &0     \\
 $5$    &0.0476&0.0476&0.0476&0.0476&0.0476&0.0476&0.0476&0.0476&0.019 \\
 $6$    &0     &0     &0     & 0    &0     &0     &0     &0     &0     \\
 $7$    &0     &0     &0     & 0    &0     &0     &0     &0     &0     \\
\hline
\end{tabular}}
%\end{table}
\vspace{3mm}
\renewcommand{\arraystretch}{1.05}
%\begin{table}[t]
\centering\caption{Cache placement matrix for $K=7$,
$N=9$, $\theta=1.5$, $M=7$.}\label{solutionM7}\resizebox{8.75cm}{!}{
\begin{tabular}{|p{1mm}<{\centering}|p{6mm}<{\centering}|p{6mm}<{\centering}|p{6mm}<{\centering}|p{6mm}<{\centering}|p{6mm}<{\centering}|p{6mm}<{\centering}|p{6mm}<{\centering}|p{6mm}<{\centering}|p{6mm}<{\centering}|p{6mm}<{\centering}} %{|c|c|c|c|c|c|c|c|c|c|c}
\hline
\multirow{2}{*}{$\abf$}&\multicolumn{9}{c|}{Cache placement vector of each file}\\ \cline{2-10}
          &$\abf_1$&$\abf_2$&$\abf_3$&$\abf_4$&$\abf_5$&$\abf_6$&$\abf_7$&$\abf_8$&$\abf_9$\\
          \hline\hline
 $0$    &0     &0     &0     &0     &0     &0     &0     &0     &0     \\
 $1$    &0     &0     &0     & 0    &0     &0     &0     &0     &0     \\
 $2$    &0     &0     &0     &0     &0     &0     &0     &0     &0     \\
 $3$    &0     &0     &0     &0     &0     &0     &0     &0     &0     \\
 $4$    &0     &0     &0     &0     &0     &0     &0     &0     &0     \\
 $5$    &0.0265&0.0265&0.0265&0.0265&0.0265&0.0265&0.0265&0.0265&0.0265\\
 $6$    &0.0635&0.0635&0.0635&0.0635&0.0635&0.0635&0.0635&0.0635&0.0635\\
 $7$    &0     &0     &0     & 0    &0     &0     &0     &0     &0     \\
\hline
\end{tabular}}
\end{table}

\subsection{Performance of Average Rate}
\begin{figure}
 \centering
  \includegraphics[scale=0.50]{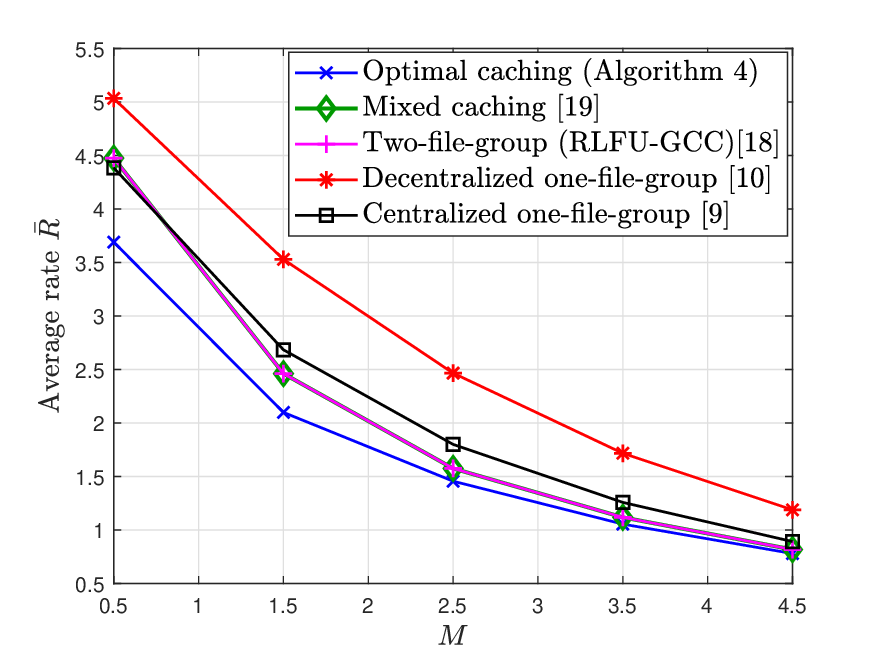}
   \caption{Average rate $\bar{R}$ vs. cache size $M$ ($N=10$, $K=6$, Zipf  distribution: $\theta=1.5$).}
  \label{fig:perfoamance_1}
%\end{figure}
%\begin{figure}
 \centering
  \includegraphics[scale=0.50]{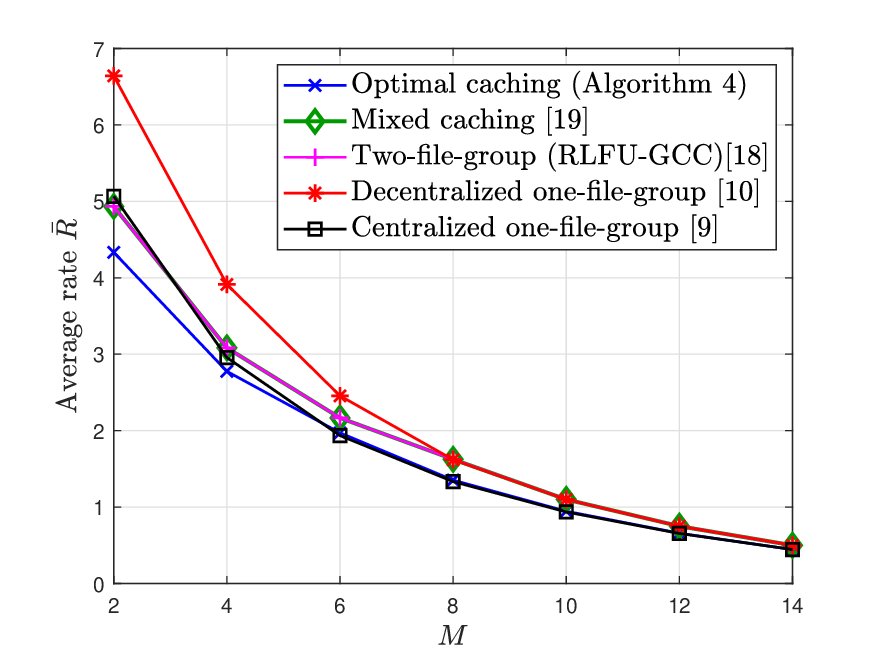}
  \caption{Average rate $\bar{R}$ vs. cache size $M$  ($N=21$, $K=12$, step-function  distribution).}
  \label{fig:perfoamance_2}
%\vspace*{-1.2em}
\end{figure}

To evaluate the performance of the  optimal cache placement scheme obtained by Algorithm~\ref{alg:placement}, we plot the average rate $\bar{R}$ vs. $M$   for file popularity using Zipf distribution and a step function in Figs.~\ref{fig:perfoamance_1} and \ref{fig:perfoamance_2}, respectively.   For comparison, we consider the centralized  \cite{Maddah-Ali&Niesen:TIT2014} and decentralized  \cite{Niesen&Maddah-Ali:TIT2015}  symmetric cache placement  schemes designed for  uniform file popularity (\ie one file group), the RLFU-GCC scheme with two file groups~\cite{Ji&Order:TIT17}, and the mixed  caching scheme in~\cite{Zhang&Coded:TIT18}. In  Fig.~\ref{fig:perfoamance_1},  we set $N=10$, $K=6$, and  Zipf parameter $\theta=1.5$. The optimal cache placement   by Algorithm~\ref{alg:placement} results in the lowest $\bar{R}$ among all the schemes. As expected, the fixed one-file-group scheme, designed for uniform popularity, has the worst performance.  The two-file-group scheme (RLFU-GCC) and the mixed caching scheme have almost identical performance. The performance gap between the two-file-group scheme (RLFU-GCC) and the optimal solution is more noticeable for smaller $M$, and reduces as $M$ increases.

In Fig.~\ref{fig:perfoamance_2}, we consider a case studied in \cite{Zhang&Coded:TIT18} with $N=21$, $K=12$, and a non-Zipf  step-function file popularity distribution given as: $p_1=5/9$, $p_n=1/30$, for $n=2,\ldots, 11$, and $p_n=1/90$, for $n=12,\ldots, 21$.   Again, the average rate under the  optimal cache placement is lower than that of all other schemes, with the gap more noticeable for smaller $M$. As an example, for $M=2$, the optimal  $\{\abf_n\}$ results in three file groups for coded caching that has  not been considered in any existing scheme.

%%%%%%%%%%%%%%%%%%%%%%%%%%%%%%%%%%%%%%%%%%%%%%%%%%%%%%
\subsection{Converse Bound}
\begin{figure}[t]
%\psfrag{$F$}{$L$}
 \centering
  \includegraphics[scale=0.50]{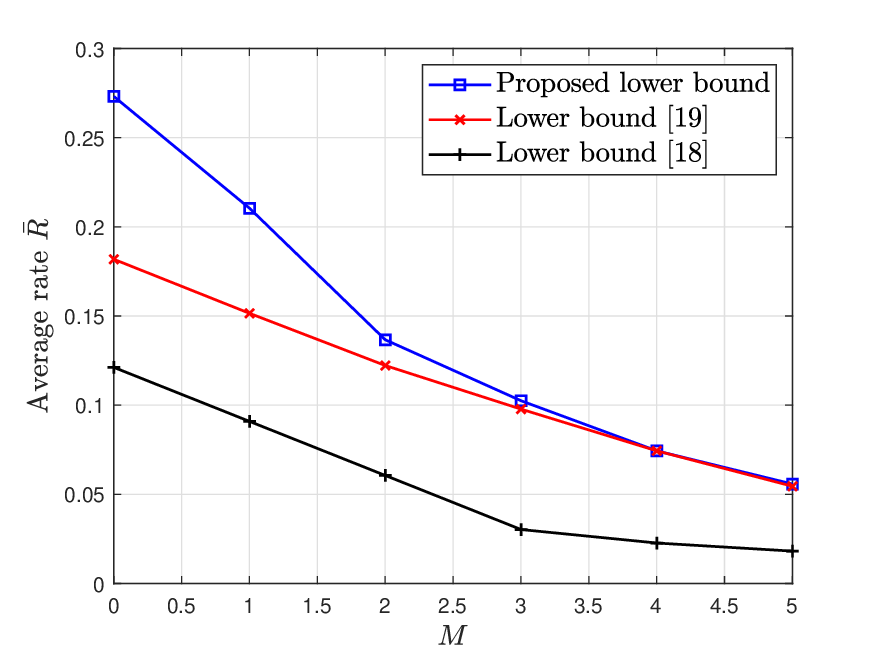}
  \caption{The lower bound  $\bar{R}^\textrm{lb}$ vs. cache size $M$ ($N=10$, $K=6$, Zipf  distribution: $\theta=1.5$ ).}
  \label{fig:lower_bound_p}
\end{figure}

We now compare our proposed lower bound in \eqref{equ_lb_opt} with those proposed in~\cite{Ji&Order:TIT17} and \cite{Zhang&Coded:TIT18}, as well as  the average rate under the optimal caching scheme  by Algorithm~\ref{alg:placement}. In Fig.~\ref{fig:lower_bound_p}, we set $N=10$, $K=6$, and Zipf parameter $\theta=1.5$. We observe that our proposed lower bound  is the highest for all values of $M$, and the gap is larger for smaller $M$. In particular, our  bound in \eqref{equ_lb_opt} based on  the optimal file groups in the cache placement is higher than the one in \eqref{equ_lowerbound_zhang} from \cite{Zhang&Coded:TIT18}.

As discussed in Section~\ref{sec:bound}, the difference in the lower bounds comes from how the values of $p',N_{p'},N_{p'}^m$  in \eqref{equ_lowerbound_merge} are set by each scheme (\ie \eqref{equ_lowerbound_zhang} and \eqref{equ_lb_opt}). To see the difference between our scheme  and two other schemes in    \cite{Zhang&Coded:TIT18}, including the two-file-group-based method and the exhaustive search, we show the values of $N_{p'}$, $N_{p'}^m$, and $\bar{R}^\textrm{lb}$ in \eqref{equ_lowerbound_merge} for each scheme in Table~\ref{table:lower_bound}. We consider Zipf distribution with $\theta=1.5$,   $K=6$, $M=1$, and compare the performance  for $N=5,7,9$.
%Different number of files leads to different file popularity vector $\pbf$.
%For $N=5$, file popularity vector $\pbf=[0.568,0.2008,0.1093,0.071,0.0508]$;
%For $N=7$, $\pbf=[0.5312,0.1878,0.1022,0.0664,0.0475,0.0361,0.0287]$; and %for $N=9$, $\pbf=[0.509,0.18,0.098, 0.064,0.046,0.035]$, $[0.027,0.022,0.019]$. \tcr{here.}
Again, our scheme  always leads to the tightest lower bound $\bar R^{\textrm{lb}}$. 
Note that for the popular file group, the optimal cache placement in our scheme always gives smaller $N_{p^{\prime}}$. This indicates that a smaller number of the most popular files are selected, in contrast to the two-file-group based method and the exhaustive search.
This shows that  even the exhaustive search used in \eqref{equ_lowerbound_zhang} is not enough to find the optimal number of the most popular files since it only searches a subgroup of the possible cases.

\begin{table}[h]
\renewcommand\arraystretch{1.2}
\centering
\caption{Comparison of  different schemes to compute the lower bound in \eqref{equ_lowerbound_merge} ($M=1$, $K=6$, Zipf distribution: $\theta=1.5$.
For $N=5$: $\pbf=[0.57,0.2,0.11,0.07,0.05]$.
For $N=7$: $\pbf=[0.53,0.19,0.1,0.07,0.05,0.04,0.03]$. For $N=9$: $\pbf=[0.51,0.18,0.1,0.06,0.05,0.04,0.03,0.02,0.02]$).
}
\begin{tabular}{|c|c|c|c|c|}
\hline
\multicolumn{2}{|c|}{\multirow{2}{*}{}} &
\multirow{2}{*}{\tabincell{c}{Two-file-\\group \cite{Zhang&Coded:TIT18}}} & \multirow{2}{*}{\tabincell{c}{Exhaustive \\Search \cite{Zhang&Coded:TIT18}}}&
\multirow{2}{*}{\tabincell{c}{Proposed}} \\
%\cline{2-3}
 \multicolumn{2}{|c|}{}& \multicolumn{1}{c|}{} &\multicolumn{1}{c|}{}&\multicolumn{1}{c|}{}
\\
\hline\hline
\multirow{3}{*}{$N=5$}
& $N_{p^{\prime}}$ & 4 & 5 & 3
\\
\cline{2-5}
& $N_{p^{\prime}}^{m}$ & 0 & 0 & 1
\\
\cline{2-5}
& ${\bar R}^{\textrm{lb}}$ & 0.0909 & 0.1109 & 0.1789
\\
\hline\hline

\multirow{3}{*}{$N=7$}
& $N_{p^{\prime}}$ & 4 & 5 & 3
\\
\cline{2-5}
& $N_{p^{\prime}}^{m}$ & 1 & 1 & 1
\\
\cline{2-5}
& ${\bar R}^{\textrm{lb}}$ & 0.1212 & 0.1296 & 0.1673
\\
\hline\hline

\multirow{3}{*}{$N=9$}
& $N_{p^{\prime}}$ & 4 & 5 & 3
\\
\cline{2-5}
& $N_{p^{\prime}}^{m}$ & 2 & 1 & 2
\\
\cline{2-5}
& ${\bar R}^{\textrm{lb}}$ & 0.1515 & 0.1242 & 0.2138
\\
\hline
\end{tabular}\label{table:lower_bound}
\end{table}

%%%%%%%%%%%%%%%%%%%%%%%%%%%%%%%%%%%%%%%
\subsection{Subpacketization Level}\label{sim:Subpkt}

\begin{figure}[t]
%\psfrag{$F$}{$L$}
 \centering
  \includegraphics[scale=0.5]{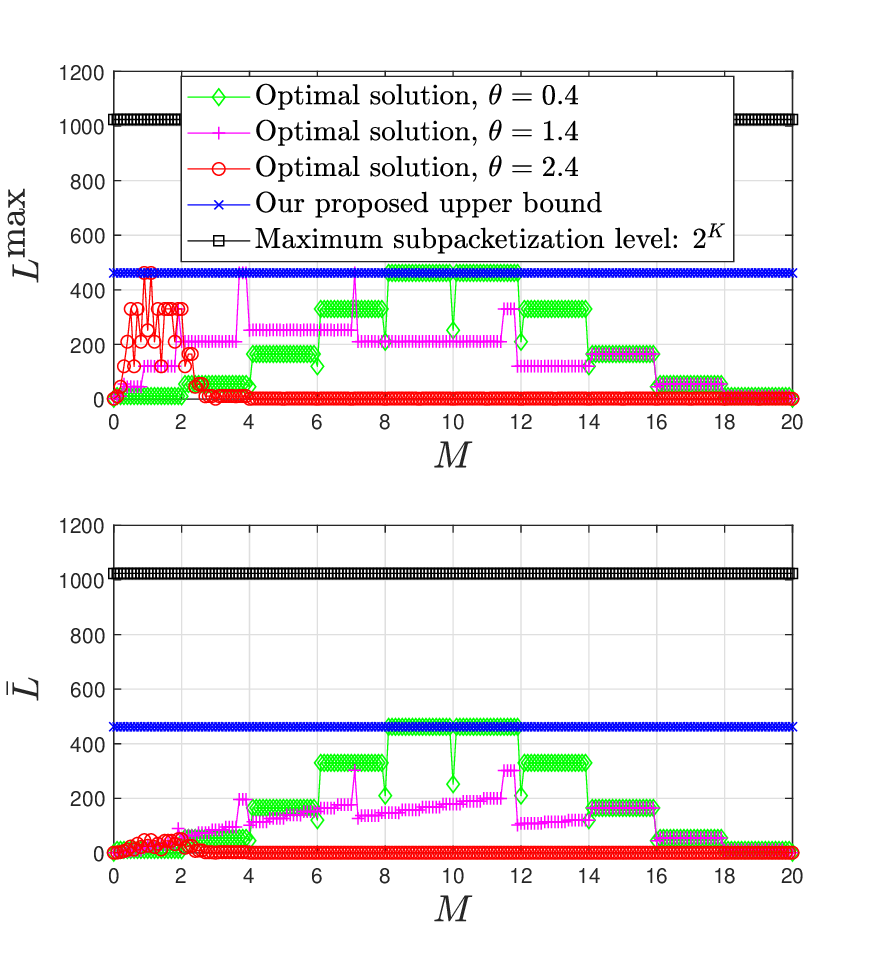}\vspace*{-1em}
  \caption{The subpacketization level under the optimal cache placement  vs. cache size $M$ ( $N=20$, $K=10$). Top: The worst-case subpacketization level $L^{\max}$. Bottom: The average subpacketization level $\bar L$.  }
  \label{fig:subpkt_pall_Zipf}
\end{figure}

Define the average subpacketization level among $N$ files by $\bar{L}=\frac{1}{N}\sum_nL_n$. For $N=20$  and $K=10$, we obtain both  $L^{\max}$ and  $\bar L$ under the optimal cache placement by solving \textbf{P1} for different $M$ and $\theta$.
Note that smaller $\theta$ indicates a more uniform popularity distribution and vice versa.  Fig.~\ref{fig:subpkt_pall_Zipf} Top shows  $L^{\max}$, the worst-case level (the upper bound in \eqref{L_bd}), and the maximum possible number of subfiles ($2^K$), over different $M$, for  Zipf parameter $\theta=0.4, 1.4, 2.4$.
We see that except for a small range of $M$, for most of the values of $M$,   $L^{\max}$ is much lower than the worst-case level.
% For $\theta=2.4$,  $L^{\max}$ reaches the upper bound when $M$ is around $2$ and decreases quickly to a very low level for $M>3$.
% For $\theta=1.4$ and $0.4$, $L^{\max}$ is the highest when $M$ is around $10$, and reduces quickly for $M$ being either smaller or larger than $10$.
Fig.~\ref{fig:subpkt_pall_Zipf} Bottom shows $\bar L$ over $M$. The general trend is similar to that of $L^{\max}$, except that $\bar L$  can be much less than $L^{\max}$ at the lower range of $M$, especially for $\theta=2.4$, where there are only a few highly popular files. For both $L^{\max}$ and $\bar{L}$, they tend to increase then decrease with $M$. This is because the location $l_o$ of the nonzero element  in $\abf_n$ increases as $M$ becomes larger, as seen  in Tables~\ref{solutionM1}--\ref{solutionM6}. As a result, the number of subfiles $\binom{K}{l_o}$ increases then decreases. In general, the subpacketization level is low  for smaller or larger $M/N$ and higher for moderate $M/N$.

\section{Conclusion}\label{sec:conclusion}
In this work,  we obtained the optimal uncoded cache placement  solution for  the CCS under arbitrary file popularity distribution and thoroughly characterized the solution structure. We identified the inherent file group structure under the optimal placement. There are at most three file groups under the optimal solution, regardless of file popularity or other system parameters. For each possible file group structure, we obtained the cache placement solution  in closed-form. Following this, we developed a simple and efficient algorithm to obtain the optimal cache placement solution by comparing a set of candidate closed-form solutions  computed in parallel.
Our insight into at most three  file groups links the caching strategy to ``most popular," ``moderately popular," and ``non-popular" file categories. Furthermore, the optimal cache placement may explore  coding opportunities across file groups for the maximum caching gain. Using our optimal cache placement, we provided a new converse bound on the average delivery rate of any coded caching scheme tighter than any existing ones.
The optimal solution structure allows us to quantify the subpacketization level under the optimal cache placement, where we showed that the worst-case subpacketization level grows as $\Oc(2^K/\sqrt{K})$.
The simulation study verified the file group structure and the optimal cache placement solution obtained by our proposed simple algorithm. The performance achieved by the optimal cache placement   was also demonstrated.

{Note that the optimal cache placement  is obtained in a centralized scenario, where the solution and the determination of file groups requires the knowledge of $K$. The knowledge of $K$ is also required for the existing file grouping strategies \cite{Ji&Order:TIT17,Zhang&Coded:TIT18} for a bounded performance. In practice, the system can estimate $K$  if it is unknown.  A careful estimation of  $K$ based on some prior information will enable us to directly apply the optimal cache placement solution obtained in this work.  For the effect of $K$ on the cache placement, in general, the cache placement for the CCS is related to the ratio  $KM/N$. A larger $K$ value means higher  $KM/N$. This leads to  more files being shifted from the ``non-popular" group to the ``most popular" group and stored in the user caches, as discussed in Section~\ref{sec:discussion}.
 The inaccurate knowledge of $K$ may result in a mismatch  to the optimal file groups and a loss from the optimal performance. Quantifying the effect of overestimating or underestimating $K$ on the performance   loss is non-trivial and needs further study as a future work. To this end, it would be also interesting to study the optimal cache placement design and its gap to the lower bound for unknown $K$ under nonuniform file popularity. 
}

\appendices
\section{Probability Distribution of $Y_m$}\label{ProbExpression}
For $Y_m$ being the $m$th smallest file index in the demand vector $\dbf$, $m=1,\ldots,K$,  the probability distribution of $Y_m$ is given as follows  \cite[Lemma 2]{Daniel&Yu:TIT19}:
\begin{align}
\Pr[Y_1=i]=&\Big(\sum_{l=i}^{N}p_l\Big)^K-\Big( \sum_{l=i+1}^{N}p_l\Big)^K,\nn\\
\Pr[Y_2=i]=&\Pr[Y_1=i]+K \Big[ \Big(\sum_{l=1}^{i-1}p_l\Big)\Big(\sum_{l=i}^{N}p_l\Big)^{K-1}\Big.\nn\\
&\Big. -\Big(\sum_{l=1}^{i}p_l\Big)\Big(\sum_{l=i+1}^{N}p_l\Big)^{K-1} \Big];\nn
\end{align}
For $3 \le m \le K$,
\begin{align}
&\quad\Pr[Y_m=1]=\sum_{k=0}^{K-m}\!\binom{K}{m+k}p_1^{m+k}(1-p_1)^{K-m-k}, \nn\\
&\quad\Pr[Y_m=i]=\nn\\
&\binom{K}{K-m+1} \left(\Big(\sum_{l=i}^{N}p_l\Big)^{K-m+1}\!\!\!\!-\Big(\sum_{l=i+1}^{N}p_l\Big)^{K-m+1}\right)\nn\\
&\left(\sum_{l=1}^{i-1}\right)^{m-1}+\sum_{k=0}^{K-2}\sum_{b=\max\{0,m-2-k\}}^{\min\{m,K-k\}-2}\nn\\
&\!\!\left(\!\!\frac{K!p_i^{2+k}}{(2+k)!b!(K-2-k-b)!}\left(\sum_{l=1}^{i-1}p_l\!\!\right)^b\!\!\!\left(\sum_{l=i+1}^{N}p_l\!\right)^{K\!-2-k-b}\!\right)\!, \quad\nn\\  
&\quad\quad\quad\quad\quad\quad\quad\quad\quad\quad\quad\quad\quad\quad\quad\quad\quad\text{for} \ i=2,\ldots,N.\nn
\end{align}

\section{Proof of Theorem~\ref{The3Groups}}\label{Proof:The3Groups}
\IEEEproof
We prove Theorem  \ref{The3Groups} by exploring the properties in the KKT conditions for \textbf{P2}.
%We first introduce the Lagrange multipliers $\gammabf_n=[\gamma_{n,1},\ldots,\gamma_{n,K}]^T$ for \eqref{ConstraintPopFir}, $\rhobf=[\rho_1,\ldots,\rho_K]^T$ for \eqref{Constraint_a_Nl}, $\rho_0$ for \eqref{Constraint_a_10}, $\nubf=[\nu_{1},\ldots,\nu_N]^{T}$ for \eqref{Constraint_SumTo1} and $\lambda$ for \eqref{Constraint_SumLeM} respectively, and derive the Lagrangian function of \textrm{\bf P2} as:
The Lagrangian associate with \textrm{\bf P2} is given by
\begin{align}
%L&(\mathbf{a_{n}}, \gammabf_n, \rhobf, \rho_0, \nubf, \lambda)=\nn\\
L&=\sum_{n=1}^{N}\gbf_n^{T}\abf_n-\sum_{n=1}^{N-1}\sum_{l=1}^{K}\gamma_{n,l}(a_{n,l}-a_{n+1,l})-\sum_{l=1}^{K}\rho_l a_{N,l}\nn\\
&-\rho_0 a_{1,0}+\lambda(\sum_{n=1}^{N}\cbf^T\abf_n - M)+\sum_{n=1}^{N}\nu_n(\bbf^T\abf_n-1)\label{Lagarangian}
\end{align}
where   $\{\gamma_{n,l}\}$ are the  Lagrange multipliers for constraint \eqref{ConstraintPopFir}, $\{\rho_0,\ldots,\rho_K\}$ are the  Lagrange multipliers for constraints in \eqref{Constraint_a_Nl}, $\{\nu_n\}$ are the  Lagrange multipliers for constraint \eqref{Constraint_SumTo1}, and $\lambda$ is  the  Lagrange multiplier for constraint \eqref{Constraint2}.
Since \textrm{\bf P2} is an LP, the KKT conditions hold for \textrm{\bf P2}, which are listed below:
\begin{align}
  &\bbf^T\abf_n=1, \label{KKTcon1} \\
  &\sum_{n=1}^{N}\cbf^T\abf_n= M, \label{KKTcon2}\\
  &a_{N,l}\geq0,\quad l\in\Kc, \label{KKTcon3}\\
  &\rho_l\cdot a_{N,l}=0, \rho_l\geq0,\quad l\in\Kc, \label{KKTcon4}\\
  &a_{1,0}\geq0, \label{KKTcon5}\\
  &\rho_0\cdot a_{1,0}=0, \label{KKTcon6}\\
  &a_{n,l}-a_{n+1,l}\geq0 ,\quad n\in\Nc\backslash\{N\},l\in\Kc,\label{KKTcon7}\\
  &\gamma_{n,l}(a_{n,l}\!-\!a_{n+1,l})\!=\!0,\gamma_{n,l}\geq0,\quad n\in\Nc\backslash\{N\},l\in\Kc \label{KKTcon8}\\
%\end{align}
%
%By taking the derivatives of $a_{n,l},n\in\Nc, l\in\Kc\bigcup0$, for $l\in\Kc$, we have:
%\begin{align}
  &\frac{\partial L}{\partial{a_{n,l}}}\!=\!g_{n,l}\!-\!\gamma_{n,l}\!+\!\gamma_{n-1,l}\!+\!\lambda c_l\!+\!\nu_n b_l=0,\nn\\
  &\hspace*{13em}n\in\Nc\backslash\{1,N\},l\in\Kc\label{KKTcon_anl}\\
  &\frac{\partial L}{\partial{a_{1,l}}}=g_{1,l}-\gamma_{1,l}+\lambda c_l+\nu_1 b_l=0,\quad l\in\Kc\label{KKTcon_a1l}\\
  &\frac{\partial L}{\partial{a_{N,l}}}=g_{N,l}-\rho_l+\gamma_{N-1,l}+\lambda c_l+\nu_N b_l=0, \label{KKTcon_aNl}\\
%\end{align}
%
%For $l=0$, we have:
%\begin{align}
&\frac{\partial L}{\partial{a_{n,0}}}={g_{n,0}}+\lambda c_0+\nu_n b_0=0,\quad n\in\Nc\backslash\{1\},\label{KKTcon_an0}\\
&\frac{\partial L}{\partial{a_{1,0}}}={g_{1,0}}-\rho_0+\lambda c_0+\nu_1 b_0=0.\label{KKTcon_a10}
\end{align}

From \eqref{KKTcon_anl} and \eqref{KKTcon_a1l}, we have, for $m=1,\ldots,N-1$,
\begin{align}
\sum_{n=1}^{m}\frac{\partial L}{\partial{a_{n,l}}}=\sum_{n=1}^{m}g_{n,l}-\gamma_{m,l}+m\lambda
c_l+\sum_{n=1}^{m}\nu_n b_l=0.\label{KKT_summ_1}
\end{align}

Based on the above KKT conditions, we prove Theorem~\ref{The3Groups} by contradiction. Assume that there exists an optimal solution $\{\abf_n\}$ that divides the files into four file groups. The structure
of the sub-placement vectors $\bar\abf_n$'s can be expressed as $\bar\abf_{1}=\ldots= \bar\abf_{n_o}\succcurlyeq_1 \bar\abf_{n_o+1}=\ldots= \bar\abf_{n_1}\succcurlyeq_1 \bar\abf_{n_1+1}=\ldots=\bar\abf_{n_2}\succcurlyeq_1 \bar\abf_{n_2+1}=\ldots=\bar\abf_{N}$, for $1\le n_o<n_1<n_2\le N-1$.
By the property in \eqref{ConstraintPopFir}, we assume $a_{n_o,l_o}> a_{n_o+1,l_o}$, $a_{n_1,l_1}> a_{n_1+1,l_1}$ and $a_{n_2,l_2}> a_{n_2+1,l_2}$, for some $l_o,l_1,l_2 \in \Kc$. From \eqref{KKTcon8}, we have
\begin{align}
\gamma_{n_o,l_o}=\gamma_{n_1,l_1}=\gamma_{n_2,l_2}=0.\label{equ:gamma0}
\end{align}
Since $c_0=0$ and $b_0=1$, from \eqref{KKTcon_an0}, we have\begin{equation}
\nu_n=-g_{n,0},\quad n\in\Nc\backslash\{1\}. \label{KKTcon_nv_n}
\end{equation}
Following \eqref{KKT_summ_1}, let $m=n_o,n_1,n_2$, we have
\begin{align}
\sum_{n=1}^{n_o}g_{n,l_o}-\gamma_{n_o,l_o}+n_o\lambda c_{l_o}+\sum_{n=1}^{n_o}\nu_n b_{l_o}=0,\label{ThreeGroups_1}\\
\sum_{n=1}^{n_1}g_{n,l_1}-\gamma_{n_1,l_1}+n_1\lambda c_{l_1}+\sum_{n=1}^{n_1}\nu_n b_{l_1}=0,\label{ThreeGroups_2}\\
\sum_{n=1}^{n_2}g_{n,l_2}-\gamma_{n_2,l_2}+n_2\lambda c_{l_2}+\sum_{n=1}^{n_2}\nu_n b_{l_2}=0.\label{ThreeGroups_3}
\end{align}
Substituting the values of $\gamma_{n_o,l_o}$, $\gamma_{n_1,l_1}$, $\gamma_{n_2,l_2}$  in \eqref{equ:gamma0} and $\nu_n$ in \eqref{KKTcon_nv_n} into  \eqref{ThreeGroups_1} - \eqref{ThreeGroups_3}, we have
\begin{align}
\lambda n_o c_{l_o}+\nu_1b_{l_o}=-\sum_{i=2}^{n_o}g_{n,0}b_{l_o}-\sum_{n=1}^{n_o}g_{n,l_o},\label{ThreeGroups_5}\\
\lambda n_1 c_{l_1}+\nu_1b_{l_1}=-\sum_{i=2}^{n_1}g_{n,0}b_{l_1}-\sum_{n=1}^{n_1}g_{n,l_1},\label{ThreeGroups_6}\\
\lambda n_2 c_{l_2}+\nu_1b_{l_2}=-\sum_{i=2}^{n_2}g_{n,0}b_{l_2}-\sum_{n=1}^{n_2}g_{n,l_2}.\label{ThreeGroups_7}
\end{align}

We  rewrite \eqref{ThreeGroups_5} - \eqref{ThreeGroups_7} into a matrix form $\Abf \xbf = \bbf$ as
\begin{align}
\hspace*{-.5em}\begin{bmatrix}
n_o c_{l_o} & b_{l_o}\\ %& \sum_{i=2}^{n_o}g_{n,0}b_{l_o}-\sum_{n=1}^{n_o}g_{n,l_o}\\
n_1 c_{l_1} & b_{l_1}\\ %& \sum_{i=2}^{n_1}g_{n,0}b_{l_1}-\sum_{n=1}^{n_1}g_{n,l_1}\\
n_2 c_{l_2} & b_{l_2} %& \sum_{i=2}^{n_2}g_{n,0}b_{l_2}-\sum_{n=1}^{n_2}g_{n,l_2}
\end{bmatrix}
\begin{bmatrix}
\lambda \\
\nu_1
\end{bmatrix}
\!=\!
\begin{bmatrix}
-\sum_{n=2}^{n_o}g_{n,0}b_{l_o}-\sum_{n=1}^{n_o}g_{n,l_o} \\
-\sum_{n=2}^{n_1}g_{n,0}b_{l_1}-\sum_{n=1}^{n_1}g_{n,l_1} \\
-\sum_{n=2}^{n_2}g_{n,0}b_{l_2}-\sum_{n=1}^{n_2}g_{n,l_2}
\end{bmatrix}.\label{TwoGroupOneDiffEle}
\end{align}
Note that $n_o \neq n_1 \neq n_2$, and by the definition of $c_l$ and $b_l$ below \eqref{ValueC_nl},  the $3\times 2$ coefficient matrix $\Abf$ in \eqref{TwoGroupOneDiffEle} is full rank, there is no feasible solution for $\lambda,\nu_1$. This contradicts the assumption that there exists an optimal $\{\abf_n\}$ with four file groups. A similar argument follows to show   more than four file groups is not possible. Thus, we have the conclusion in Theorem~\ref{The3Groups}.
\endIEEEproof

\section{Proof of Proposition~\ref{Pro:TwoGroupOneDiffEle1}}\label{ProofPro:TwoGroupOneDiffEle1}
\IEEEproof
With two file groups, the sub-placement vectors have the following relation: $\bar\abf_1=\ldots=\bar\abf_{n_o}\succcurlyeq_1\bar\abf_{n_o+1}=\ldots=\bar\abf_{N}$, for some $ n_o \in \{1,\ldots, N-1\}$.
Since there is at least one  element that is different between $\bar\abf_{n_o}$ and $\bar\abf_{n_o+1}$, by \eqref{ConstraintPopFir}, we assume $a_{n_o,l_o}>a_{n_o+1,l_o}$, for some $l_o \in \Kc$. Consequently, we have $\gamma_{n_o,l_o}=0$  based on   \eqref{KKTcon8}.
From \eqref{KKT_summ_1}, we have
\begin{align}
  \sum_{n=1}^{n_o}g_{n,l}+n_o\lambda c_{l_o}+\sum_{n=1}^{n_o}\nu_n b_{l_o}=0.\label{TwoGroupOneDiffEle3}
\end{align}

 From \eqref{KKTcon_anl}--\eqref{KKTcon_aNl}, we have
 \begin{align}
   \sum_{n=1}^{N}g_{n,l}-\rho_{l}+N\lambda c_{l}+\sum_{n=1}^{N}\nu_n b_{l}=0,\quad l\in\Kc.      \label{KKT_summ_4}
 \end{align}

%We present our proof by contradiction.
Assume $\bar\abf_{N}$ has two nonzero elements at the $l_1$th and $l_2$th locations, \ie $a_{N,l_1}> 0$, $a_{N,l_2}> 0$, for $l_1\neq l_2$, $l_1,l_2\in \Kc$. Note that one of $l_1$ and $l_2$ can be $l_o$. Without loss of generality, we assume $l_2\neq l_o$. We know from \eqref{KKTcon4} that $\rho_{l_1}=\rho_{l_2}=0$.
Then, from \eqref{KKT_summ_4}, we have
\begin{align}
  \sum_{n=1}^{N}g_{n,l_1}+N\lambda c_{l_1}+\sum_{n=1}^{N}\nu_n b_{l_1}=0, \label{TwoGroupOneDiffEle1}\\
  \sum_{n=1}^{N}g_{n,l_2}+N\lambda c_{l_2}+\sum_{n=1}^{N}\nu_n b_{l_2}=0.\label{TwoGroupOneDiffEle2}
\end{align}
Using the expression of $\nu_n$ in \eqref{KKTcon_nv_n}, we can rewrite \eqref{TwoGroupOneDiffEle3}\eqref{TwoGroupOneDiffEle1}\eqref{TwoGroupOneDiffEle2} into a matrix form as
\begin{align}
\hspace*{-.5em}\begin{bmatrix}
n_o c_{l_o} & b_{l_o}\\ %& \sum_{i=2}^{n_o}g_{n,0}b_{l_o}-\sum_{n=1}^{n_o}g_{n,l_o}\\
N c_{l_1} & b_{l_1}\\ %& \sum_{i=2}^{n_1}g_{n,0}b_{l_1}-\sum_{n=1}^{n_1}g_{n,l_1}\\
N c_{l_2} & b_{l_2} %& \sum_{i=2}^{n_2}g_{n,0}b_{l_2}-\sum_{n=1}^{n_2}g_{n,l_2}
\end{bmatrix}
\begin{bmatrix}
\lambda \\
\nu_1
\end{bmatrix}
\!=\!
\begin{bmatrix}
-\sum_{n=2}^{n_o}g_{n,0}b_{l_o}-\sum_{n=1}^{n_o}g_{n,l_o} \\
-\sum_{n=2}^{N}g_{n,0}b_{l_1}-\sum_{n=1}^{N}g_{n,l_1} \\
-\sum_{n=2}^{N}g_{n,0}b_{l_2}-\sum_{n=1}^{N}g_{n,l_2}
\end{bmatrix}.\label{ThreeFileGroup}
\end{align}

Similar to the argument in the proof of Theorem~\ref{The3Groups},
 since $n_o<N$, $l_2 \neq l_1$, $l_2\neq l_o$,  by the definition of $c_l$ and $b_l$, the coefficient matrix of \eqref{ThreeFileGroup} is full rank,  and there is no feasible solution for $\lambda$ and $\nu_1$, contradicting the assumption that the optimal $\bar\abf_N$ has two nonzero elements. Similarly, we show the optimal $\bar\abf_N$ cannot have more than two nonzero elements. Thus, we complete the proof.
\endIEEEproof

%%%%%%%%%%%%%%%%%%%%%%%%%%%%
\section{Proof of Proposition \ref{Pro:TwoGroupsTwoUnequals}}\label{ProofPro:TwoGroupsTwoUnequals}
\IEEEproof
Since the optimal cache placement solution result in  two file groups, the sub-placement vectors have the following structure: $\bar\abf_{1}= \ldots= \bar\abf_{n_o}\succcurlyeq_1 \bar\abf_{n_o+1}=\ldots= \bar\abf_{N}\succcurlyeq_1 \bf 0$, for some $n_o\in \{1,\ldots,N-1\}$.
By Proposition \ref{Pro:TwoGroupOneDiffEle1},  $\bar{\abf}_{n_o+1}$  has only one nonzero element. Assume $a_{n_o+1,l_o}>0$, for some $l_o\in \Kc$. From \eqref{KKTcon4}, we know that $\rho_{l_o}=0$.
Then from \eqref{KKT_summ_4}, we have
\begin{align}\label{OneUneq6}
  \sum_{n=1}^{N}g_{n,l_o}+\lambda N c_{l_o}+\sum_{n=1}^{N}\nu_nb_{l_o}=0.
\end{align}

Assume that there are two  elements in $\bar\abf_{n_o}$ and $\bar\abf_{n_o+1}$ being different: $a_{n_o,l_1}>a_{n_o+1,l_1}$ and $a_{n_o,l_2}>a_{n_o+1,l_2}$, for $l_1\neq l_2$,  $l_1,l_2\in \Kc$. Without loss of generality, we assume $l_2\neq l_o$. From \eqref{KKTcon8}, we have $\gamma_{n_o,l_{1}}=\gamma_{n_o,l_{2}}=0$. As a result, from \eqref{KKT_summ_1}, we have
\begin{align}
\sum_{n=1}^{n_o}g_{n,l_{1}}+\lambda n_o c_{l_{1}}+\sum_{n=1}^{n_o}\nu_n b_{l_{1}}=0,\label{OneUneq1}\\
\sum_{n=1}^{n_o}g_{n,l_{2}}+\lambda n_oc_{l_{2}}+\sum_{n=1}^{n_o}\nu_nb_{l_{2}}=0.\label{OneUneq2}
\end{align}
Again, using \eqref{KKTcon_nv_n}, we put \eqref{OneUneq6}--\eqref{OneUneq2}  in a matrix form as
 \begin{align}
\begin{bmatrix}
N c_{l_o} & b_{l_o}\\ %& \sum_{i=2}^{n_o}g_{n,0}b_{l_o}-\sum_{n=1}^{n_o}g_{n,l_o}\\
n_o c_{l_1} & b_{l_1}\\ %& \sum_{i=2}^{n_1}g_{n,0}b_{l_1}-\sum_{n=1}^{n_1}g_{n,l_1}\\
n_o c_{l_2} & b_{l_2} %& \sum_{i=2}^{n_2}g_{n,0}b_{l_2}-\sum_{n=1}^{n_2}g_{n,l_2}
\end{bmatrix}
\begin{bmatrix}
\lambda \\
\nu_1
\end{bmatrix}
=
\begin{bmatrix}
-\sum_{n=2}^{N}g_{n,0}b_{l_o}\!-\!\sum_{n=1}^{N}g_{n,l_o} \\
-\sum_{n=2}^{n_o}g_{n,0}b_{l_1}\!-\!\sum_{n=1}^{n_o}g_{n,l_1} \\
-\sum_{n=2}^{n_o}g_{n,0}b_{l_2}\!-\!\sum_{n=1}^{n_o}g_{n,l_2}
\end{bmatrix}.\nn%\label{TwoGroupOneDiffEle1}
\end{align}
By the similar argument in the proof of Proposition~\ref{Pro:TwoGroupOneDiffEle1}, the coefficient matrix of \eqref{OneUneq6}--\eqref{OneUneq2} is full rank, and  $\lambda$ and $\nu_1$ do not have any feasible solution, contradicting the assumption that $\bar\abf_N$ has two nonzero elements. Similarly, we can proof that  $\bar\abf_N$ cannot have more than two nonzero elements. \endIEEEproof

\section{Proof of Proposition \ref{Pro:TwoGroupsFirstZero}}\label{ProofPro:TwoGroupsFirstZero}
\IEEEproof
With two file groups, the sub-placement vectors have the following structure: $\bar\abf_{1}= \ldots= \bar\abf_{n_o}\succcurlyeq_1 \bar\abf_{n_o+1}=\ldots= \bar\abf_{N}$, for some $n_o\in \{1,\ldots,N-1\}$. Assume $\bar a_{n_o+1,l_o}>0$, for $l_o\in \Kc$. By Proposition \ref{Pro:TwoGroupsTwoUnequals}, only one element is different between $\bar{\abf}_{n_o}$ and $\bar\abf_{n_o+1}$. This element can be either at $l_o$, \ie $a_{n_o,l_o}>a_{n_o+1,l_o}$ (as shown in Fig.~\ref{fig:TwoGroup2}), or any $ l_1\neq l_o$, $l_1\in \Kc$, \ie $a_{n_o,l_1}>a_{n_o+1,l_1}$ (as shown in Fig.~\ref{fig:TwoGroup3}). We discuss the two cases separately.

\subsubsection{$a_{n_o,l_o}>a_{n_o+1,l_o}>0$}
If $a_{1,0}=\ldots=a_{n_o,0}>0$, by \eqref{KKTcon6} we have $\rho_0=0$. Combining this with \eqref{KKTcon_a10}, we have ${g_{1,0}}+\lambda c_0+\nu_1 b_0=0$, which gives
\begin{align}\label{OneDiffOneNonZero1}
\nu_1=-g_{1,0}.
\end{align}
By  \eqref{KKTcon8}, since $a_{n_o,l_o}>a_{n_o+1,l_o}$, we have $\gamma_{n_o,l_o}=0$. Substituting the expression of  $\nu_n$  in  \eqref{KKTcon_nv_n} and \eqref{OneDiffOneNonZero1} into \eqref{KKT_summ_1}, we have
\begin{align}\label{OneDiffOneNonZero2}
  \lambda n_o c_{l_o}=-\sum_{n=1}^{n_o}g_{n,{l_o}}-\sum_{n=1}^{n_o}g_{n,0} b_{l_o}.
\end{align}

 Combining (\ref{KKTcon_anl}) and \eqref{KKTcon_aNl}, we have
 \begin{align}\label{KKT_summ_2}
&\!\!\sum_{n=n_o+1}^{N}\frac{\partial L}{\partial{a_{n,l_o}}} \nn\\
 &=\sum_{n=n_o+1}^{N}\!\!g_{n,l_o}-\rho_{l_o}+\!\gamma_{n_o,l_o}+(N-n_o)\lambda c_{l_o}+\!\!\sum_{n=n_o+1}^{N}\!\!\nu_n b_{l_o} \nn\\
 &= 0.
 \end{align}

For $a_{n_o+1,l_o}=a_{N,l_o}>0$, from \eqref{KKTcon4}, we have $\rho_{l_o}=0$. Along with $\gamma_{n_o,l_o}=0$,  \eqref{KKT_summ_2} can be rewritten as
\begin{align}\label{OneDiffOneNonZero3}
\lambda(N-n_o) c_{l_o}=-\!\!\!\!\sum_{n=n_o+1}^{N}\!\!g_{n,l_o}-\!\!\!\!\sum_{n=n_o+1}^{N}\!\!g_{n,0}b_{l_o}.
\end{align}

Examining \eqref{OneDiffOneNonZero2} and \eqref{OneDiffOneNonZero3}, we see that there is no feasible solution for $\lambda$ to satisfy both equations. This contradicts  the assumption that $a_{1,0}=\ldots=a_{n_o,0}> 0$.
%%%%%%%%
\subsubsection{$a_{n_o,l_1}>a_{n_o+1,l_1}=0$, for $l_1\neq l_o$}
From  \eqref{KKTcon8}, we have $\gamma_{n_o,l_1}=0$. Assuming $a_{1,0}=\ldots=a_{n_o,0}>0$, we have \eqref{OneDiffOneNonZero1}. Similar to \eqref{OneDiffOneNonZero2}, we have
\begin{align}\label{OneDiffOneNonZero4}
\lambda n_o c_{l_1}=-\sum_{n=1}^{n_o}g_{n,{l_1}}-\sum_{n=1}^{n_o}g_{n,0} b_{l_1}.
\end{align}
For $a_{n_o+1,l_o}=a_{N,l_o}>0$, again we have $\rho_{l_o}=0$, and $\nu_n$
  in  \eqref{KKTcon_nv_n} and \eqref{OneDiffOneNonZero1}. Thus, from  \eqref{KKT_summ_4}, we have
\begin{align}\label{OneDiffOneNonZero5}
\lambda N c_{l_o} =-\sum_{n=2}^{N}g_{n,0}b_{l_o}\!-\!\sum_{n=1}^{N}g_{n,l_o}.
\end{align}
Again, there is no feasible solution for $\lambda$ to satisfy both \eqref{OneDiffOneNonZero4} and \eqref{OneDiffOneNonZero5}.
       This contradicts the assumption that $a_{1,0}=\ldots=a_{n_o,0}>0$.

From both two cases above, we conclude  if $a_{1,0}=\ldots=a_{n_o,0}>0$, there is no feasible solution for  $\lambda$ and $\nu_1$. Thus, we have $a_{1,0}=\ldots=a_{n_o,0}=0$ for the optimal $\{\abf_n\}$.
\endIEEEproof

%%%%%%%%%%%%%%%%%%%%%%%%%%%%
\section{Proof of Proposition \ref{Pro:ThirdZero}}\label{ProofPro:ThirdZero}
\IEEEproof
With three file groups, the sub-placement vectors have the following structure:  $\bar\abf_{1}=\ldots= \bar\abf_{n_o}\succcurlyeq_1 \bar\abf_{n_o+1}=\ldots= \bar\abf_{n_1}\succcurlyeq_1 \bar\abf_{n_1+1}=\ldots=\bar\abf_{N}$, for  $1\le n_o<n_1\le N-1$. Assume that $a_{n_o,l_o}>a_{n_o+1,l_o}$ and $a_{n_1,l_1}>a_{n_1+1,l_1}$, for $l_o,l_1 \in \Kc$. From \eqref{KKTcon8}, we have $\gamma_{n_o,l_o}=\gamma_{n_1,l_1}=0$. Substitute the value of $\nu_n$ in \eqref{KKTcon_nv_n} into  \eqref{KKT_summ_1}, we have the following
\begin{align}
\lambda n_o c_{l_o}+\nu_1b_{l_o}=-\sum_{n=2}^{n_o}g_{n,0}b_{l_o}-\sum_{n=1}^{n_o}g_{n,l_o},\label{ThreeGroups_51}\\
\lambda n_1 c_{l_1}+\nu_1b_{l_1}=-\sum_{n=2}^{n_1}g_{n,0}b_{l_1}-\sum_{n=1}^{n_1}g_{n,l_1}\label{ThreeGroups_61}.
\end{align}
To show $\bar\abf_{n_1+1}=\mathbf{0}$ by contradiction, assume that $\bar{\abf}_{N}\succcurlyeq_1\mathbf{0}$, \ie $\bar\abf_{n_1+1}=\ldots=\bar\abf_{N}$ has at least one nonzero element.
Let $a_{N,l_2}>0$ for some $l_2\in \Kc$. Then, we have $\rho_{l_2}=0$ by \eqref{KKTcon4}. Then, from \eqref{KKT_summ_4}, we have
\begin{align}\label{ThreeGroupsThird3}
  N\lambda c_{l_2}+\nu_1b_{l_2}=-\sum_{n=1}^{N}g_{n,l_2}-\sum_{n=2}^{N}g_{n,0} b_{l_2}.
\end{align}
Putting \eqref{ThreeGroups_51}--\eqref{ThreeGroupsThird3} into a matrix form, we have
 \begin{align}
\begin{bmatrix}
 n_o c_{l_o} & b_{l_o} \\ %& \sum_{n=2}^{n_o}g_{n,0}b_{l_o}-\sum_{n=1}^{n_o}g_{n,l_o}\\
 n_1 c_{l_1} & b_{l_1} \\ %& \sum_{n=2}^{n_1}g_{n,0}b_{l_1}-\sum_{n=1}^{n_1}g_{n,l_1}\\
 N c_{l_2} & b_{l_2}  %& \sum_{n=2}^{n_2}g_{n,0}b_{l_2}-\sum_{n=1}^{n_2}g_{n,l_2}
\end{bmatrix}
 \cdot
\begin{bmatrix}
 \lambda  \\
 \nu_1
\end{bmatrix}
 =
\begin{bmatrix}
 -\sum_{n=2}^{n_o}g_{n,0}b_{l_o} - \sum_{n=1}^{n_o}g_{n,l_o}  \\
 -\sum_{n=2}^{n_1}g_{n,0}b_{l_1} - \sum_{n=1}^{n_1}g_{n,l_1}  \\
 -\sum_{n=2}^{N}g_{n,0}b_{l_2} - \sum_{n=1}^{N}g_{n,l_2}
\end{bmatrix}. \nn %\label{Third0}
\end{align}
Using a similar argument as in the proof of Theorem~\ref{The3Groups}, we  conclude
that $\lambda$ and $\nu_1$ do not have any feasible solution, which contradicts the assumption that $\bar\abf_N\succcurlyeq_1 {\bf 0}$. Thus, we complete the proof.
\endIEEEproof

\bibliographystyle{IEEEtran}
\bibliography{IEEEabrv,Yong}
\end{document}